\documentclass[longauth]{aa}
\usepackage{euclid}

\usepackage{graphicx}
\usepackage{natbib}
\usepackage{ulem}
\usepackage{scalerel}
\usepackage{comment}
\usepackage[table]{xcolor}

\usepackage[pdfencoding=auto,psdextra]{hyperref}
\hypersetup{
    colorlinks=true,
    linkcolor=blue,
    urlcolor=blue,
    citecolor=blue
}
\usepackage{url} 
\makeatletter
\renewcommand*\aa@pageof{, page \thepage{} of \pageref*{LastPage}}
\makeatother

\usepackage[utf8]{inputenc}

\usepackage{newtxtext}
\usepackage{siunitx}

\begin{document}

   \title{\Euclid: Measuring the intrinsic alignment of galaxies around cosmic voids in the \Euclid Flagship simulation\thanks{This paper is published on behalf of the Euclid Consortium.}}    

%%%% Version Wednesday 10th of June 2026 02:58:07 PM UT
%%%% Assumes the new A&A style file from Oct 2025 or later
%%%% Please do not edit the author list -- contact ECEB Bureau for changes
%%%% Version Thursday 2nd of July 2026 09:20:28 AM UT
%%%% Assumes the new A&A style file from Oct 2025 or later
%%%% Please do not edit the author list -- contact ECEB Bureau for changes
\newcommand{\orcid}[1]{} %% if already defined in aa.cls: comment, or use renewcommand			   
\author{P.~Vielzeuf\orcid{0000-0003-2035-9339}\thanks{\email{vielzeuf@cppm.in2p3.fr}}\inst{\ref{aff1}}
\and M.-C.~Cousinou\inst{\ref{aff1}}
\and A.~Pisani\orcid{0000-0002-6146-4437}\inst{\ref{aff1}}
\and S.~Escoffier\orcid{0000-0002-2847-7498}\inst{\ref{aff1}}
\and E.~J.~Gonzalez\orcid{0000-0002-0226-9893}\inst{\ref{aff2},\ref{aff3}}
\and W.~d'Assignies~\orcid{0000-0002-9719-1717}\inst{\ref{aff4}}
\and E.~Jullo\orcid{0000-0002-9253-053X}\inst{\ref{aff5}}
\and M.~Ghodsi~Yengejeh\orcid{0000-0001-5481-9810}\inst{\ref{aff6},\ref{aff7},\ref{aff8}}
\and A.~Kov\'acs\orcid{0000-0002-5825-579X}\inst{\ref{aff6},\ref{aff7}}
\and S.~Andreon\orcid{0000-0002-2041-8784}\inst{\ref{aff9}}
\and N.~Auricchio\orcid{0000-0003-4444-8651}\inst{\ref{aff10}}
\and C.~Baccigalupi\orcid{0000-0002-8211-1630}\inst{\ref{aff11},\ref{aff12},\ref{aff13},\ref{aff14}}
\and M.~Baldi\orcid{0000-0003-4145-1943}\inst{\ref{aff15},\ref{aff10},\ref{aff16}}
\and S.~Bardelli\orcid{0000-0002-8900-0298}\inst{\ref{aff10}}
\and P.~Battaglia\orcid{0000-0002-7337-5909}\inst{\ref{aff10}}
\and A.~Biviano\orcid{0000-0002-0857-0732}\inst{\ref{aff12},\ref{aff11}}
\and E.~Branchini\orcid{0000-0002-0808-6908}\inst{\ref{aff17},\ref{aff18},\ref{aff9}}
\and M.~Brescia\orcid{0000-0001-9506-5680}\inst{\ref{aff19},\ref{aff20}}
\and S.~Camera\orcid{0000-0003-3399-3574}\inst{\ref{aff21},\ref{aff22},\ref{aff23}}
\and V.~Capobianco\orcid{0000-0002-3309-7692}\inst{\ref{aff23}}
\and C.~Carbone\orcid{0000-0003-0125-3563}\inst{\ref{aff24}}
\and V.~F.~Cardone\inst{\ref{aff25},\ref{aff26}}
\and J.~Carretero\orcid{0000-0002-3130-0204}\inst{\ref{aff27},\ref{aff28}}
\and M.~Castellano\orcid{0000-0001-9875-8263}\inst{\ref{aff25}}
\and G.~Castignani\orcid{0000-0001-6831-0687}\inst{\ref{aff10}}
\and S.~Cavuoti\orcid{0000-0002-3787-4196}\inst{\ref{aff20},\ref{aff29}}
\and A.~Cimatti\inst{\ref{aff30}}
\and C.~Colodro-Conde\inst{\ref{aff31}}
\and G.~Congedo\orcid{0000-0003-2508-0046}\inst{\ref{aff32}}
\and L.~Conversi\orcid{0000-0002-6710-8476}\inst{\ref{aff33},\ref{aff34}}
\and Y.~Copin\orcid{0000-0002-5317-7518}\inst{\ref{aff35}}
\and F.~Courbin\orcid{0000-0003-0758-6510}\inst{\ref{aff36},\ref{aff37},\ref{aff38}}
\and H.~M.~Courtois\orcid{0000-0003-0509-1776}\inst{\ref{aff39}}
\and H.~Degaudenzi\orcid{0000-0002-5887-6799}\inst{\ref{aff40}}
\and S.~de~la~Torre\inst{\ref{aff5}}
\and G.~De~Lucia\orcid{0000-0002-6220-9104}\inst{\ref{aff12}}
\and H.~Dole\orcid{0000-0002-9767-3839}\inst{\ref{aff41}}
\and F.~Dubath\orcid{0000-0002-6533-2810}\inst{\ref{aff40}}
\and X.~Dupac\inst{\ref{aff34}}
\and M.~Farina\orcid{0000-0002-3089-7846}\inst{\ref{aff42}}
\and R.~Farinelli\inst{\ref{aff10}}
\and S.~Farrens\orcid{0000-0002-9594-9387}\inst{\ref{aff43}}
\and F.~Faustini\orcid{0000-0001-6274-5145}\inst{\ref{aff25},\ref{aff44}}
\and S.~Ferriol\inst{\ref{aff35}}
\and F.~Finelli\orcid{0000-0002-6694-3269}\inst{\ref{aff10},\ref{aff45}}
\and P.~Fosalba\orcid{0000-0002-1510-5214}\inst{\ref{aff46},\ref{aff47}}
\and S.~Fotopoulou\orcid{0000-0002-9686-254X}\inst{\ref{aff48}}
\and N.~Fourmanoit\orcid{0009-0005-6816-6925}\inst{\ref{aff1}}
\and M.~Frailis\orcid{0000-0002-7400-2135}\inst{\ref{aff12}}
\and E.~Franceschi\orcid{0000-0002-0585-6591}\inst{\ref{aff10}}
\and M.~Fumana\orcid{0000-0001-6787-5950}\inst{\ref{aff24}}
\and S.~Galeotta\orcid{0000-0002-3748-5115}\inst{\ref{aff12}}
\and K.~George\orcid{0000-0002-1734-8455}\inst{\ref{aff49}}
\and W.~Gillard\orcid{0000-0003-4744-9748}\inst{\ref{aff1}}
\and B.~Gillis\orcid{0000-0002-4478-1270}\inst{\ref{aff32}}
\and C.~Giocoli\orcid{0000-0002-9590-7961}\inst{\ref{aff10},\ref{aff16}}
\and P.~G\'omez-Alvarez\orcid{0000-0002-8594-5358}\inst{\ref{aff50},\ref{aff34}}
\and J.~Gracia-Carpio\orcid{0000-0003-4689-3134}\inst{\ref{aff51}}
\and A.~Grazian\orcid{0000-0002-5688-0663}\inst{\ref{aff52}}
\and F.~Grupp\inst{\ref{aff51},\ref{aff53}}
\and S.~V.~H.~Haugan\orcid{0000-0001-9648-7260}\inst{\ref{aff54}}
\and H.~Hoekstra\orcid{0000-0002-0641-3231}\inst{\ref{aff55}}
\and W.~Holmes\inst{\ref{aff56}}
\and F.~Hormuth\inst{\ref{aff57}}
\and A.~Hornstrup\orcid{0000-0002-3363-0936}\inst{\ref{aff58},\ref{aff59}}
\and K.~Jahnke\orcid{0000-0003-3804-2137}\inst{\ref{aff60}}
\and M.~Jhabvala\inst{\ref{aff61}}
\and B.~Joachimi\orcid{0000-0001-7494-1303}\inst{\ref{aff62}}
\and S.~Kermiche\orcid{0000-0002-0302-5735}\inst{\ref{aff1}}
\and A.~Kiessling\orcid{0000-0002-2590-1273}\inst{\ref{aff56}}
\and M.~Kilbinger\orcid{0000-0001-9513-7138}\inst{\ref{aff43}}
\and B.~Kubik\orcid{0009-0006-5823-4880}\inst{\ref{aff35}}
\and M.~K\"ummel\orcid{0000-0003-2791-2117}\inst{\ref{aff53}}
\and M.~Kunz\orcid{0000-0002-3052-7394}\inst{\ref{aff63}}
\and H.~Kurki-Suonio\orcid{0000-0002-4618-3063}\inst{\ref{aff64},\ref{aff65}}
\and A.~M.~C.~Le~Brun\orcid{0000-0002-0936-4594}\inst{\ref{aff66}}
\and S.~Ligori\orcid{0000-0003-4172-4606}\inst{\ref{aff23}}
\and P.~B.~Lilje\orcid{0000-0003-4324-7794}\inst{\ref{aff54}}
\and V.~Lindholm\orcid{0000-0003-2317-5471}\inst{\ref{aff64},\ref{aff65}}
\and I.~Lloro\orcid{0000-0001-5966-1434}\inst{\ref{aff67}}
\and G.~Mainetti\orcid{0000-0003-2384-2377}\inst{\ref{aff68}}
\and O.~Mansutti\orcid{0000-0001-5758-4658}\inst{\ref{aff12}}
\and O.~Marggraf\orcid{0000-0001-7242-3852}\inst{\ref{aff69}}
\and M.~Martinelli\orcid{0000-0002-6943-7732}\inst{\ref{aff25},\ref{aff26}}
\and N.~Martinet\orcid{0000-0003-2786-7790}\inst{\ref{aff5}}
\and F.~Marulli\orcid{0000-0002-8850-0303}\inst{\ref{aff70},\ref{aff10},\ref{aff16}}
\and R.~J.~Massey\orcid{0000-0002-6085-3780}\inst{\ref{aff71}}
\and S.~Maurogordato\inst{\ref{aff72}}
\and E.~Medinaceli\orcid{0000-0002-4040-7783}\inst{\ref{aff10}}
\and S.~Mei\orcid{0000-0002-2849-559X}\inst{\ref{aff73},\ref{aff74}}
\and M.~Meneghetti\orcid{0000-0003-1225-7084}\inst{\ref{aff10},\ref{aff16}}
\and E.~Merlin\orcid{0000-0001-6870-8900}\inst{\ref{aff25}}
\and G.~Meylan\inst{\ref{aff75}}
\and A.~Mora\orcid{0000-0002-1922-8529}\inst{\ref{aff76}}
\and M.~Moresco\orcid{0000-0002-7616-7136}\inst{\ref{aff70},\ref{aff10}}
\and L.~Moscardini\orcid{0000-0002-3473-6716}\inst{\ref{aff70},\ref{aff10},\ref{aff16}}
\and C.~Neissner\orcid{0000-0001-8524-4968}\inst{\ref{aff4},\ref{aff28}}
\and S.-M.~Niemi\orcid{0009-0005-0247-0086}\inst{\ref{aff77}}
\and J.~W.~Nightingale\orcid{0000-0002-8987-7401}\inst{\ref{aff78}}
\and C.~Padilla\orcid{0000-0001-7951-0166}\inst{\ref{aff4}}
\and S.~Paltani\orcid{0000-0002-8108-9179}\inst{\ref{aff40}}
\and F.~Pasian\orcid{0000-0002-4869-3227}\inst{\ref{aff12}}
\and K.~Pedersen\inst{\ref{aff79}}
\and V.~Pettorino\orcid{0000-0002-4203-9320}\inst{\ref{aff77}}
\and A.~Pezzotta\orcid{0000-0003-0726-2268}\inst{\ref{aff9}}
\and S.~Pires\orcid{0000-0002-0249-2104}\inst{\ref{aff43}}
\and G.~Polenta\orcid{0000-0003-4067-9196}\inst{\ref{aff44}}
\and L.~A.~Popa\inst{\ref{aff80}}
\and L.~Pozzetti\orcid{0000-0001-7085-0412}\inst{\ref{aff10}}
\and F.~Raison\orcid{0000-0002-7819-6918}\inst{\ref{aff51}}
\and A.~Renzi\orcid{0000-0001-9856-1970}\inst{\ref{aff81},\ref{aff82},\ref{aff10}}
\and J.~Rhodes\orcid{0000-0002-4485-8549}\inst{\ref{aff56}}
\and G.~Riccio\inst{\ref{aff20}}
\and E.~Romelli\orcid{0000-0003-3069-9222}\inst{\ref{aff12}}
\and M.~Roncarelli\orcid{0000-0001-9587-7822}\inst{\ref{aff10}}
\and C.~Rosset\orcid{0000-0003-0286-2192}\inst{\ref{aff73}}
\and R.~Saglia\orcid{0000-0003-0378-7032}\inst{\ref{aff53},\ref{aff51}}
\and Z.~Sakr\orcid{0000-0002-4823-3757}\inst{\ref{aff83},\ref{aff84},\ref{aff85}}
\and A.~G.~S\'anchez\orcid{0000-0003-1198-831X}\inst{\ref{aff51}}
\and D.~Sapone\orcid{0000-0001-7089-4503}\inst{\ref{aff86}}
\and B.~Sartoris\orcid{0000-0003-1337-5269}\inst{\ref{aff53},\ref{aff12}}
\and P.~Schneider\orcid{0000-0001-8561-2679}\inst{\ref{aff69}}
\and T.~Schrabback\orcid{0000-0002-6987-7834}\inst{\ref{aff87}}
\and A.~Secroun\orcid{0000-0003-0505-3710}\inst{\ref{aff1}}
\and E.~Sihvola\orcid{0000-0003-1804-7715}\inst{\ref{aff88}}
\and P.~Simon\inst{\ref{aff69}}
\and C.~Sirignano\orcid{0000-0002-0995-7146}\inst{\ref{aff81},\ref{aff82}}
\and G.~Sirri\orcid{0000-0003-2626-2853}\inst{\ref{aff16}}
\and A.~Spurio~Mancini\orcid{0000-0001-5698-0990}\inst{\ref{aff89}}
\and L.~Stanco\orcid{0000-0002-9706-5104}\inst{\ref{aff82}}
\and P.~Tallada-Cresp\'{i}\orcid{0000-0002-1336-8328}\inst{\ref{aff27},\ref{aff28}}
\and A.~N.~Taylor\inst{\ref{aff32}}
\and H.~I.~Teplitz\orcid{0000-0002-7064-5424}\inst{\ref{aff90}}
\and I.~Tereno\orcid{0000-0002-4537-6218}\inst{\ref{aff91},\ref{aff92}}
\and N.~Tessore\orcid{0000-0002-9696-7931}\inst{\ref{aff93}}
\and S.~Toft\orcid{0000-0003-3631-7176}\inst{\ref{aff94},\ref{aff95}}
\and R.~Toledo-Moreo\orcid{0000-0002-2997-4859}\inst{\ref{aff96}}
\and F.~Torradeflot\orcid{0000-0003-1160-1517}\inst{\ref{aff28},\ref{aff27}}
\and I.~Tutusaus\orcid{0000-0002-3199-0399}\inst{\ref{aff47},\ref{aff46},\ref{aff84}}
\and J.~Valiviita\orcid{0000-0001-6225-3693}\inst{\ref{aff64},\ref{aff65}}
\and T.~Vassallo\orcid{0000-0001-6512-6358}\inst{\ref{aff12},\ref{aff49}}
\and G.~Verdoes~Kleijn\orcid{0000-0001-5803-2580}\inst{\ref{aff97}}
\and A.~Veropalumbo\orcid{0000-0003-2387-1194}\inst{\ref{aff9},\ref{aff18},\ref{aff17}}
\and Y.~Wang\orcid{0000-0002-4749-2984}\inst{\ref{aff98}}
\and J.~Weller\orcid{0000-0002-8282-2010}\inst{\ref{aff53},\ref{aff51}}
\and G.~Zamorani\orcid{0000-0002-2318-301X}\inst{\ref{aff10}}
\and F.~M.~Zerbi\orcid{0000-0002-9996-973X}\inst{\ref{aff9}}
\and E.~Zucca\orcid{0000-0002-5845-8132}\inst{\ref{aff10}}
\and T.~Castro\orcid{0000-0002-6292-3228}\inst{\ref{aff12},\ref{aff13},\ref{aff11},\ref{aff99}}
\and M.~Sereno\orcid{0000-0003-0302-0325}\inst{\ref{aff10},\ref{aff16}}
\and M.~Viel\orcid{0000-0002-2642-5707}\inst{\ref{aff11},\ref{aff12},\ref{aff14},\ref{aff13},\ref{aff99}}
\and D.~Navarro-Giron\'{e}s\orcid{0000-0003-0507-372X}\inst{\ref{aff55}}}
										   
%%%% please do not edit the affiliation list -- contact ECEB Bureau for changes
\institute{Aix-Marseille Universit\'e, CNRS/IN2P3, CPPM, Marseille, France\label{aff1}
\and
Departament de F\'{\i}sica, Universitat Aut\`onoma de Barcelona, 08193 Bellaterra (Barcelona), Spain\label{aff2}
\and
Instituto de Astronomia Teorica y Experimental (IATE-CONICET), Laprida 854, X5000BGR, C\'ordoba, Argentina\label{aff3}
\and
Institut de F\'{i}sica d'Altes Energies (IFAE), The Barcelona Institute of Science and Technology, Campus UAB, 08193 Bellaterra (Barcelona), Spain\label{aff4}
\and
Aix-Marseille Universit\'e, CNRS, CNES, LAM, Marseille, France\label{aff5}
\and
MTA-CSFK Lend\"ulet Large-Scale Structure Research Group, Konkoly-Thege Mikl\'os \'ut 15-17, H-1121 Budapest, Hungary\label{aff6}
\and
Konkoly Observatory, HUN-REN CSFK, MTA Centre of Excellence, Budapest, Konkoly Thege Mikl\'os {\'u}t 15-17. H-1121, Hungary\label{aff7}
\and
ELTE E\"otv\"os Lor\'and University, Institute of Physics and Astronomy, P\'azm\'any P. st. 1/A, H-1171 Budapest, Hungary\label{aff8}
\and
INAF-Osservatorio Astronomico di Brera, Via Brera 28, 20122 Milano, Italy\label{aff9}
\and
INAF-Osservatorio di Astrofisica e Scienza dello Spazio di Bologna, Via Piero Gobetti 93/3, 40129 Bologna, Italy\label{aff10}
\and
IFPU, Institute for Fundamental Physics of the Universe, via Beirut 2, 34151 Trieste, Italy\label{aff11}
\and
INAF-Osservatorio Astronomico di Trieste, Via G. B. Tiepolo 11, 34143 Trieste, Italy\label{aff12}
\and
INFN, Sezione di Trieste, Via Valerio 2, 34127 Trieste TS, Italy\label{aff13}
\and
SISSA, International School for Advanced Studies, Via Bonomea 265, 34136 Trieste TS, Italy\label{aff14}
\and
Dipartimento di Fisica e Astronomia, Universit\`a di Bologna, Via Gobetti 93/2, 40129 Bologna, Italy\label{aff15}
\and
INFN-Sezione di Bologna, Viale Berti Pichat 6/2, 40127 Bologna, Italy\label{aff16}
\and
Dipartimento di Fisica, Universit\`a di Genova, Via Dodecaneso 33, 16146, Genova, Italy\label{aff17}
\and
INFN-Sezione di Genova, Via Dodecaneso 33, 16146, Genova, Italy\label{aff18}
\and
Department of Physics "E. Pancini", University Federico II, Via Cinthia 6, 80126, Napoli, Italy\label{aff19}
\and
INAF-Osservatorio Astronomico di Capodimonte, Via Moiariello 16, 80131 Napoli, Italy\label{aff20}
\and
Dipartimento di Fisica, Universit\`a degli Studi di Torino, Via P. Giuria 1, 10125 Torino, Italy\label{aff21}
\and
INFN-Sezione di Torino, Via P. Giuria 1, 10125 Torino, Italy\label{aff22}
\and
INAF-Osservatorio Astrofisico di Torino, Via Osservatorio 20, 10025 Pino Torinese (TO), Italy\label{aff23}
\and
INAF-IASF Milano, Via Alfonso Corti 12, 20133 Milano, Italy\label{aff24}
\and
INAF-Osservatorio Astronomico di Roma, Via Frascati 33, 00078 Monteporzio Catone, Italy\label{aff25}
\and
INFN-Sezione di Roma, Piazzale Aldo Moro, 2 - c/o Dipartimento di Fisica, Edificio G. Marconi, 00185 Roma, Italy\label{aff26}
\and
Centro de Investigaciones Energ\'eticas, Medioambientales y Tecnol\'ogicas (CIEMAT), Avenida Complutense 40, 28040 Madrid, Spain\label{aff27}
\and
Port d'Informaci\'{o} Cient\'{i}fica, Campus UAB, C. Albareda s/n, 08193 Bellaterra (Barcelona), Spain\label{aff28}
\and
INFN section of Naples, Via Cinthia 6, 80126, Napoli, Italy\label{aff29}
\and
Dipartimento di Fisica e Astronomia "Augusto Righi" - Alma Mater Studiorum Universit\`a di Bologna, Viale Berti Pichat 6/2, 40127 Bologna, Italy\label{aff30}
\and
Instituto de Astrof\'{\i}sica de Canarias, E-38205 La Laguna, Tenerife, Spain\label{aff31}
\and
Institute for Astronomy, University of Edinburgh, Royal Observatory, Blackford Hill, Edinburgh EH9 3HJ, UK\label{aff32}
\and
European Space Agency/ESRIN, Largo Galileo Galilei 1, 00044 Frascati, Roma, Italy\label{aff33}
\and
ESAC/ESA, Camino Bajo del Castillo, s/n., Urb. Villafranca del Castillo, 28692 Villanueva de la Ca\~nada, Madrid, Spain\label{aff34}
\and
Universit\'e Claude Bernard Lyon 1, CNRS/IN2P3, IP2I Lyon, UMR 5822, Villeurbanne, F-69100, France\label{aff35}
\and
Institut de Ci\`{e}ncies del Cosmos (ICCUB), Universitat de Barcelona (IEEC-UB), Mart\'{i} i Franqu\`{e}s 1, 08028 Barcelona, Spain\label{aff36}
\and
Instituci\'o Catalana de Recerca i Estudis Avan\c{c}ats (ICREA), Passeig de Llu\'{\i}s Companys 23, 08010 Barcelona, Spain\label{aff37}
\and
Institut de Ciencies de l'Espai (IEEC-CSIC), Campus UAB, Carrer de Can Magrans, s/n Cerdanyola del Vall\'es, 08193 Barcelona, Spain\label{aff38}
\and
UCB Lyon 1, CNRS/IN2P3, IUF, IP2I Lyon, 4 rue Enrico Fermi, 69622 Villeurbanne, France\label{aff39}
\and
Department of Astronomy, University of Geneva, ch. d'Ecogia 16, 1290 Versoix, Switzerland\label{aff40}
\and
Universit\'e Paris-Saclay, CNRS, Institut d'astrophysique spatiale, 91405, Orsay, France\label{aff41}
\and
INAF-Istituto di Astrofisica e Planetologia Spaziali, via del Fosso del Cavaliere, 100, 00100 Roma, Italy\label{aff42}
\and
Universit\'e Paris-Saclay, Universit\'e Paris Cit\'e, CEA, CNRS, AIM, 91191, Gif-sur-Yvette, France\label{aff43}
\and
Space Science Data Center, Italian Space Agency, via del Politecnico snc, 00133 Roma, Italy\label{aff44}
\and
INFN-Bologna, Via Irnerio 46, 40126 Bologna, Italy\label{aff45}
\and
Institut d'Estudis Espacials de Catalunya (IEEC),  Edifici RDIT, Campus UPC, 08860 Castelldefels, Barcelona, Spain\label{aff46}
\and
Institute of Space Sciences (ICE, CSIC), Campus UAB, Carrer de Can Magrans, s/n, 08193 Barcelona, Spain\label{aff47}
\and
School of Physics, HH Wills Physics Laboratory, University of Bristol, Tyndall Avenue, Bristol, BS8 1TL, UK\label{aff48}
\and
University Observatory, LMU Faculty of Physics, Scheinerstr.~1, 81679 Munich, Germany\label{aff49}
\and
FRACTAL S.L.N.E., calle Tulip\'an 2, Portal 13 1A, 28231, Las Rozas de Madrid, Spain\label{aff50}
\and
Max Planck Institute for Extraterrestrial Physics, Giessenbachstr. 1, 85748 Garching, Germany\label{aff51}
\and
INAF-Osservatorio Astronomico di Padova, Via dell'Osservatorio 5, 35122 Padova, Italy\label{aff52}
\and
Universit\"ats-Sternwarte M\"unchen, Fakult\"at f\"ur Physik, Ludwig-Maximilians-Universit\"at M\"unchen, Scheinerstr.~1, 81679 M\"unchen, Germany\label{aff53}
\and
Institute of Theoretical Astrophysics, University of Oslo, P.O. Box 1029 Blindern, 0315 Oslo, Norway\label{aff54}
\and
Leiden Observatory, Leiden University, Einsteinweg 55, 2333 CC Leiden, The Netherlands\label{aff55}
\and
Jet Propulsion Laboratory, California Institute of Technology, 4800 Oak Grove Drive, Pasadena, CA, 91109, USA\label{aff56}
\and
Felix Hormuth Engineering, Goethestr. 17, 69181 Leimen, Germany\label{aff57}
\and
Technical University of Denmark, Elektrovej 327, 2800 Kgs. Lyngby, Denmark\label{aff58}
\and
Cosmic Dawn Center (DAWN), Denmark\label{aff59}
\and
Max-Planck-Institut f\"ur Astronomie, K\"onigstuhl 17, 69117 Heidelberg, Germany\label{aff60}
\and
NASA Goddard Space Flight Center, Greenbelt, MD 20771, USA\label{aff61}
\and
Department of Physics and Astronomy, University College London, Gower Street, London WC1E 6BT, UK\label{aff62}
\and
Universit\'e de Gen\`eve, D\'epartement de Physique Th\'eorique and Centre for Astroparticle Physics, 24 quai Ernest-Ansermet, CH-1211 Gen\`eve 4, Switzerland\label{aff63}
\and
Department of Physics, P.O. Box 64, University of Helsinki, 00014 Helsinki, Finland\label{aff64}
\and
Helsinki Institute of Physics, Gustaf H{\"a}llstr{\"o}min katu 2, University of Helsinki, 00014 Helsinki, Finland\label{aff65}
\and
Laboratoire d'etude de l'Univers et des phenomenes eXtremes, Observatoire de Paris, Universit\'e PSL, Sorbonne Universit\'e, CNRS, 92190 Meudon, France\label{aff66}
\and
SKAO, Jodrell Bank, Lower Withington, Macclesfield SK11 9FT, UK\label{aff67}
\and
Centre de Calcul de l'IN2P3/CNRS, 21 avenue Pierre de Coubertin 69627 Villeurbanne Cedex, France\label{aff68}
\and
Universit\"at Bonn, Argelander-Institut f\"ur Astronomie, Auf dem H\"ugel 71, 53121 Bonn, Germany\label{aff69}
\and
Dipartimento di Fisica e Astronomia "Augusto Righi" - Alma Mater Studiorum Universit\`a di Bologna, via Piero Gobetti 93/2, 40129 Bologna, Italy\label{aff70}
\and
Department of Physics, Institute for Computational Cosmology, Durham University, South Road, Durham, DH1 3LE, UK\label{aff71}
\and
Universit\'e C\^{o}te d'Azur, Observatoire de la C\^{o}te d'Azur, CNRS, Laboratoire Lagrange, Bd de l'Observatoire, CS 34229, 06304 Nice cedex 4, France\label{aff72}
\and
Universit\'e Paris Cit\'e, CNRS, Astroparticule et Cosmologie, 75013 Paris, France\label{aff73}
\and
CNRS-UCB International Research Laboratory, Centre Pierre Bin\'etruy, IRL2007, CPB-IN2P3, Berkeley, USA\label{aff74}
\and
Institute of Physics, Laboratory of Astrophysics, Ecole Polytechnique F\'ed\'erale de Lausanne (EPFL), Observatoire de Sauverny, 1290 Versoix, Switzerland\label{aff75}
\and
Telespazio UK S.L. for European Space Agency (ESA), Camino bajo del Castillo, s/n, Urbanizacion Villafranca del Castillo, Villanueva de la Ca\~nada, 28692 Madrid, Spain\label{aff76}
\and
European Space Agency/ESTEC, Keplerlaan 1, 2201 AZ Noordwijk, The Netherlands\label{aff77}
\and
School of Mathematics, Statistics and Physics, Newcastle University, Herschel Building, Newcastle-upon-Tyne, NE1 7RU, UK\label{aff78}
\and
DARK, Niels Bohr Institute, University of Copenhagen, Jagtvej 155, 2200 Copenhagen, Denmark\label{aff79}
\and
Institute of Space Science, Str. Atomistilor, nr. 409 M\u{a}gurele, Ilfov, 077125, Romania\label{aff80}
\and
Dipartimento di Fisica e Astronomia "G. Galilei", Universit\`a di Padova, Via Marzolo 8, 35131 Padova, Italy\label{aff81}
\and
INFN-Padova, Via Marzolo 8, 35131 Padova, Italy\label{aff82}
\and
Instituto de F\'isica Te\'orica UAM-CSIC, Campus de Cantoblanco, 28049 Madrid, Spain\label{aff83}
\and
Institut de Recherche en Astrophysique et Plan\'etologie (IRAP), Universit\'e de Toulouse, CNRS, UPS, CNES, 14 Av. Edouard Belin, 31400 Toulouse, France\label{aff84}
\and
Universit\'e St Joseph; Faculty of Sciences, Beirut, Lebanon\label{aff85}
\and
Departamento de F\'isica, FCFM, Universidad de Chile, Blanco Encalada 2008, Santiago, Chile\label{aff86}
\and
Universit\"at Innsbruck, Institut f\"ur Astro- und Teilchenphysik, Technikerstr. 25/8, 6020 Innsbruck, Austria\label{aff87}
\and
Department of Physics and Helsinki Institute of Physics, Gustaf H\"allstr\"omin katu 2, University of Helsinki, 00014 Helsinki, Finland\label{aff88}
\and
Department of Physics, Royal Holloway, University of London, Surrey TW20 0EX, UK\label{aff89}
\and
Infrared Processing and Analysis Center, California Institute of Technology, Pasadena, CA 91125, USA\label{aff90}
\and
Departamento de F\'isica, Faculdade de Ci\^encias, Universidade de Lisboa, Edif\'icio C8, Campo Grande, PT1749-016 Lisboa, Portugal\label{aff91}
\and
Instituto de Astrof\'isica e Ci\^encias do Espa\c{c}o, Faculdade de Ci\^encias, Universidade de Lisboa, Tapada da Ajuda, 1349-018 Lisboa, Portugal\label{aff92}
\and
Mullard Space Science Laboratory, University College London, Holmbury St Mary, Dorking, Surrey RH5 6NT, UK\label{aff93}
\and
Cosmic Dawn Center (DAWN)\label{aff94}
\and
Niels Bohr Institute, University of Copenhagen, Jagtvej 128, 2200 Copenhagen, Denmark\label{aff95}
\and
Universidad Polit\'ecnica de Cartagena, Departamento de Electr\'onica y Tecnolog\'ia de Computadoras,  Plaza del Hospital 1, 30202 Cartagena, Spain\label{aff96}
\and
Kapteyn Astronomical Institute, University of Groningen, PO Box 800, 9700 AV Groningen, The Netherlands\label{aff97}
\and
Caltech/IPAC, 1200 E. California Blvd., Pasadena, CA 91125, USA\label{aff98}
\and
ICSC - Centro Nazionale di Ricerca in High Performance Computing, Big Data e Quantum Computing, Via Magnanelli 2, Bologna, Italy\label{aff99}}    
\date{\today}

% \abstract{}{}{}{}{} 
% 5 {} token are mandatory
 
\abstract {We present a methodology to measure the intrinsic alignment (IA) signal of galaxies in the vicinity of cosmic voids using the \Euclid-like Flagship cosmological simulation from the Euclid Consortium. The IA signal is quantified and compared with the predictions of the linear alignment (LA) model, providing one of the first detailed investigations of this effect in underdense large-scale environments. While the IA signal around cosmic voids has received little attention to date, it may constitute a non-negligible systematic in forthcoming cosmological analyses that exploit void-lensing measurements. Our analysis examines red and blue galaxy populations separately, enabling a comparison of their alignment behaviour in void environments with the corresponding trends measured in galaxy-galaxy correlations. We find that the redshift evolution of the IA amplitude in cosmic voids is broadly consistent with that measured in the general galaxy population for both colour-selected samples. Additionally, our modelling allows us to estimate the linear bias of voids, $b_{\rm V}(z)$ which characterises how cosmic voids trace the underlying dark-matter density field, for voids with radii in the range $10 < R_{\rm V}/(h^{-1} {\rm Mpc}) < 15$. The measured bias exhibits a positive trend with redshift, consistent with theoretical predictions for the clustering of underdense regions. These results highlight the importance of accurately modelling IA in void studies, both to mitigate systematic effects in void-lensing cosmology and to further improve our understanding of galaxy-environment interactions in low-density regions of the Universe.
}
\keywords{Surveys -- Cosmology: observations --
large-scale structure of Universe -- Gravitational lensing: weak}
   \titlerunning{\Euclid\/: Intrinsic alignment around cosmic voids}
   \authorrunning{P. Vielzeuf et al.}

\maketitle
\nolinenumbers
%
%-------------------------------------------------------------------

\section{Introduction}
Gravitational lensing has emerged as one of the most powerful probes for constraining cosmological parameters. Recent photometric surveys based on shear measurement have achieved percent-level precision in estimating the dark matter density and the amplitude of its clumpiness $\sigma_{\rm 8}$, notably through Stage III galaxy surveys such as the Dark Energy Survey (DES, \citealp{DESY3extandcosmo2023}), the Kilo-Degree Survey (KIDS, \citealp{KIDS2025}), the Hyper-Suprime Camera survey (HSC, \citealp{li2023,dalal2023}), or a combination of them \citep{DESKIDS2023}. 
However, in addition to the alignment of galaxies caused by gravitational lensing, galaxies are also influenced by tidal forces from the local gravitational field, which can intrinsically align their shapes with the surrounding gravitational potential.

This intrinsic-alignment signal acts as a contaminant in cosmic shear measurements (see, e.g., \citealp{hirata2007}); however, with appropriate modelling, it can also be exploited as a complementary cosmological probe (see, e.g. \citealp{chisari2013,Kurita,vanDompseler}, and references therein).

Indeed, because this alignment is directly correlated with the underlying local matter field, it provides a way to model the effect of the local environment on galaxy evolution and, in particular, on their orientation in space. 
However, such modelling is strongly dependent on the galaxy population under consideration. In particular, while red elliptical galaxies exhibit a relatively strong local alignment with surrounding overdensities (see, e.g.\citealp{Mandelbaum2006,hirata2007,georgiou2025,navarro2026}), no intrinsic-alignment signal has yet been detected for blue spiral galaxies \citep{hirata2007,Mandelbaum2011,Samuroff2019,Johnston2019,Samuroff2023}.

Cosmic voids, defined as large underdense regions in the cosmic web complementing the more extensively studied overdense structures such as galaxy clusters and filaments, have emerged as a promising probe for placing additional constraints on cosmological models, a prospect now corroborated by recent observations

\citep{hamaus2017,Qingqing2017,aubert2022}. Indeed, various
void statistics have proven particularly well adapted to breaking
degeneracies between cosmological parameters that typically arise when using overdense regions as cosmological probes (\citealp{sahlen2019,bayer2021,kreisch2022,contarini2022,contarini2024,bonici2023,pelliciari2023}). In particular, the lensing imprint of cosmic voids, which has already been detected in galaxy shear at the $14\,\sigma$ level (\citealp{melchior2014,clampitt,sanchez,fang}) and in  cosmic microwave background (CMB) lensing maps with up to $17\,\sigma$ confidence level \citep{vielzeuf2021,kovacs2022,camacho2024,Demirbozan2024,sartori2025}, arise from their underdense nature. This makes voids a particularly attractive probe for testing alternative cosmological models \citep{Barreira2015,baker2018,davies2019} and for investigating the presence of massive neutrinos \citep{massara2015,Kreisch2019,vielzeuf2023,Schuster2019}.

As in galaxy-galaxy lensing measurements, it is important to assess whether galaxies surrounding cosmic voids exhibit preferential orientations.
While \cite{slosar2009} reported that the orientations of spiral galaxies around voids are consistent with random orientation, \cite{trujillo2006}, \cite{varela2012}, and \cite{d'Assignies}, predicted and detected a non-zero signal (reaching up to $95.45 \%$ of the expected amplitude) for the preferential alignment of red elliptical galaxies.
Stage IV surveys, such as the \textit{Euclid} satellite \citep{mellier2025}, are expected to significantly enhance the statistical precision of weak lensing measurements.
In this context, exploiting the lensing signal of cosmic voids as a cosmological probe in the near future requires a detailed understanding of the various factors affecting galaxy orientations, including both shear and intrinsic alignments.

In this analysis, we use \Euclid-like cosmological simulations to measure and subsequently model the intrinsic alignments of galaxies residing within voids and their surroundings, using the linear alignment model proposed in \cite{d'Assignies}. 

The paper is organized as follows. In Sect.~\ref{sec:FS2}, we describe the cosmological simulation used in this analysis, along with the void catalogue derived from it. Section~\ref{sec:methodandconcept} provides a concise overview of the theoretical background relevant to gravitational lensing and intrinsic alignments. The modelling framework for void intrinsic alignments adopted in this work is detailed in Sect.~\ref{sec:model}, while Sect.~\ref{sec:estimator} introduces the statistical estimators employed to extract the alignment signal. The results of the analysis are presented and discussed in Sect.~\ref{sec:results}. Finally, Sect.~\ref{sec:conclusion} summarises the main findings and outlines prospects for future work.

\section{Flagship simulation}\label{sec:FS2}

In this work, we base our analysis on the \Euclid Flagship galaxy mock catalogue \citep{Flagship}. This galaxy catalogue was built by populating the haloes identified in the Flagship2 Wide $N$-body dark matter simulation (FS2-Wide) generated with the \texttt{PKDGRAV3} $N$-body code \citep{Potter2017} It was run assuming the \Euclid reference cosmology as defined in the \citet{Kanbenhans-EP2} Haloes were identified using the \texttt{ROCKSTAR} algorithm \citep{behroozi2013} with a minimum of 10 particles ($\sim 7.3 \times 10^{9}\, M_\odot$).

The FS2-Wide galaxy mock was produced using the `Science Pipeline at PIC', \texttt{SciPIC} \citep{Carretero2017} and consists of a light-cone covering one octant of the sky, reaching up to redshift 3. To populate the haloes, \texttt{SciPIC} combines halo occupation distribution and abundance matching techniques to assign key galaxy properties such as number density, luminosity, and colours. The parameters for this procedure are tuned to reproduce the observed galaxy clustering at low redshifts. 
Each galaxy is assigned a spectral energy distribution (SED) based on its luminosity, redshift, and colour, enabling the computation of observed fluxes across a wide range of bands by integrating the SED with the corresponding filter response. Although the mock parameters were set to reproduce the observed clustering at low redshift, the catalogue reproduces clustering measurements at higher redshifts across a wide range of galaxy populations \citep{Gonzalez2026}.
For a detailed description of the galaxy catalogue generation, we refer the reader to \cite{Flagship}.

The observed projected ellipticities of galaxies result from the combination of their intrinsic shapes and the distortions induced by gravitational lensing, given the line-of-sight mass distribution characterized by the shear-galaxy components. In FS2-Wide, the projected ellipticity components arising from shear and those from intrinsic shapes are provided separately by two distinct modules. Lensing properties, shear and convergence, are assigned following the methodology of \citet{Fosalba2008} and \citet{Fosalba:15b} by considering the whole dark matter particle distribution along the line-of-sight.

Intrinsic galaxy shapes are computed using a semi-analytical prescription, in which each galaxy is modeled as a  3-dimensional (3D) ellipsoid. First, \texttt{SciPIC} assigns two 3D axis ratios to each object so that the distribution of projected 2-dimensional (2D) axis ratios matches the observed distribution from the COSMOS survey as a function of redshift, magnitude, and colour. The 3D orientations of the galaxy principal axes are then assigned according to galaxy type, either central or satellite. Central galaxies are aligned with the principal axes of their host haloes, while satellite galaxies are oriented toward the centres of their host haloes. 
A random misalignment to these initial orientations is subsequently applied, drawn from a von Mises--Fisher distribution. The width of this distribution varies according to each galaxy’s redshift, magnitude, and colour, and is calibrated against alignment signals derived from multiple constraining data sets. Further details can be found in \citet{Hoffmann2022} and in \cite{Hoffmann}.

\subsection{Void identification and void catalogues}\label{sec:void_cat}

Recent studies have shown that 2D voids are particularly well suited for measuring the weak gravitational lensing signal of background galaxies, thanks to their well-defined underdense regions and projection effects that enhance lensing contrast (see \citealp{cautun2018, paillas2019,fang}). Building on this, we employ the void-finder algorithm developed by \cite{sanchez}, which identifies 2D underdensities within smoothed density maps constructed using the \texttt{HEALPix}\footnote{\url{https://healpix.sourceforge.io/}} pixelization scheme, where the maps are generated from the galaxy catalogue divided into slices of thickness $100\,h^{-1}\mathrm{Mpc}$. The algorithm begins by locating the most underdense pixel in the smoothed map. From this minimum, it grows a circular region by progressively including neighboring pixels until the density in the outer shell reaches the global mean of the map. The resulting region is then defined as a projected cosmic void. Within this framework, the void radius $R_{\rm V}$ is naturally defined as the scale at which the density of the surrounding shell first equals the background value, just before the onset of the compensation wall\footnote{The compensation wall refers to the overdense shell of matter that typically surrounds an underdense cosmic void.}. The procedure then iterates, moving to the next most underdense pixel not yet assigned to a previously identified void, and repeats until no significant underdensity remains.\footnote{We note that while different void-finding algorithms may adopt distinct operational definitions, such as watershed or spherical underdensity, the physical underdensities they trace are largely consistent. In practice, differences in void definition mainly influence the detailed void profiles and the resulting signal-to-noise ratio of the lensing measurement, rather than the qualitative detection of the void-lensing effect itself (see for example \citealp{paillas2019} where comparison of different algorithms is made).}

We apply this algorithm to the Flagship galaxy catalogues described in Sect. \ref{sec:FS2}, using a Gaussian smoothing scale of $5\,h^{-1}\mathrm{Mpc}$, which provides a balance between spatial resolution and noise suppression. To construct a robust void catalogue and properly account for survey boundaries, we populate the FS2 survey mask with a large number of random points distributed uniformly within the mask. For each detected void, we then compute the number density of these random points inside its volume. Voids that partially extend beyond the survey mask contain a lower density of random points and are therefore identified as boundary-affected systems. We exclude such voids from the final catalogue by applying a minimum threshold on the random-point density within the void. This post-processing step eliminates spurious edge voids that artificially extend beyond the survey footprint, thereby improving the reliability of the void sample. Additionally, we adopt the line-of-sight slicing strategy proposed by \cite{sanchez}, which enhances the precision in locating void centres along the line-of-sight and mitigates projection effects.

\begin{figure}
    \centering

    \includegraphics[width=1\linewidth]{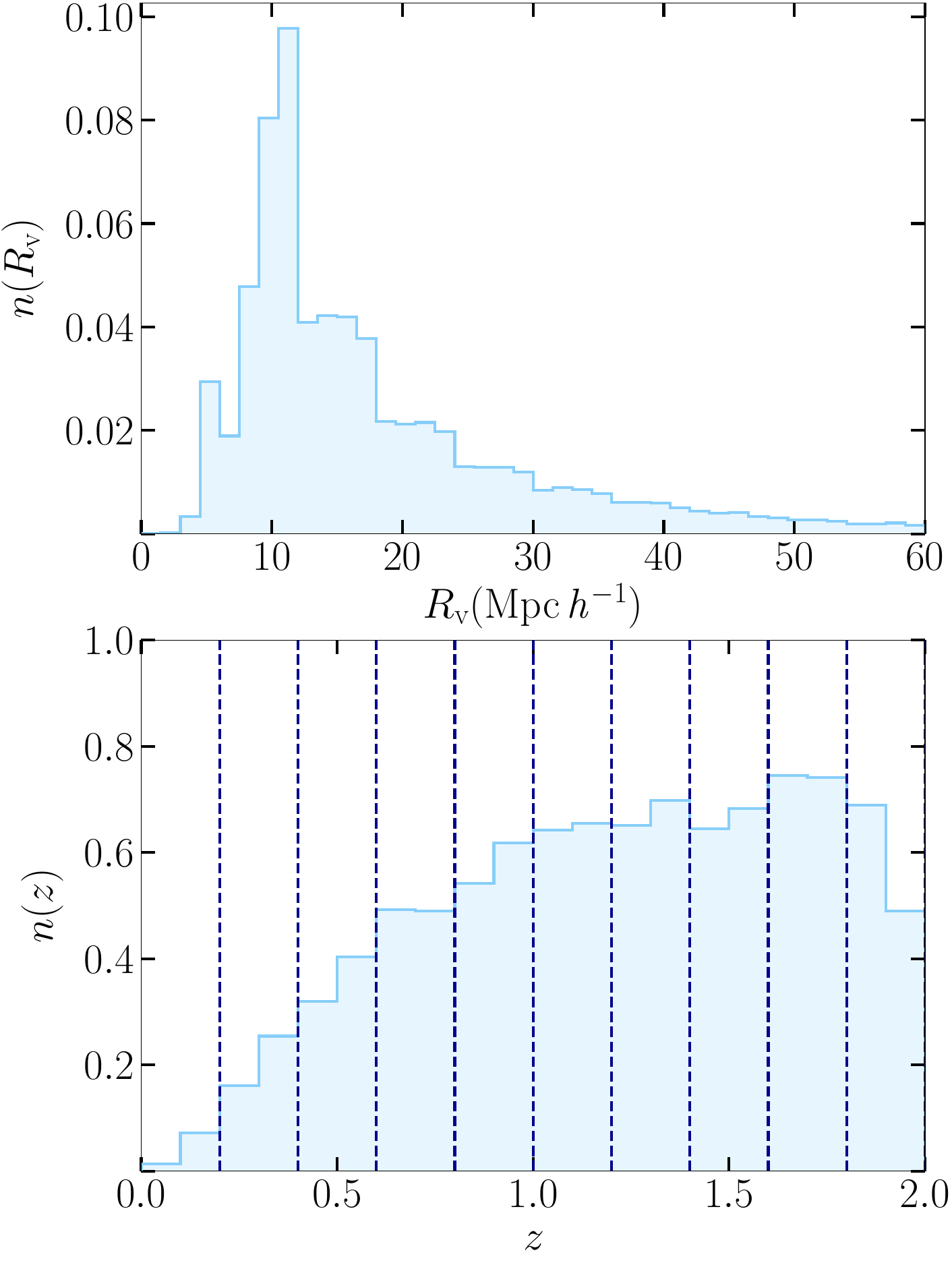}
    \caption{\textit{Top}: Normalised distribution of void radius for voids identified with the 2D void finder in the Flagship simulations. \textit{ Bottom:} Histogram of the redshift distribution of the void catalogue, with vertical dashed blue lines indicating the redshift bins used in the analysis.}
    \label{fig:voidsraddist}
\end{figure}

The void-finder algorithm is applied to the full \Euclid-like Flagship catalogue without imposing any additional cuts in colour, magnitude, or density, covering the redshift range \(0 < z < 2\), which corresponds to the interval expected to be probed by the upcoming \Euclid photometric survey. The top panel of Fig.~\ref{fig:voidsraddist} shows the normalised distribution of void radii in our catalogue, revealing a pronounced peak around $R_\mathrm{V} \sim 10\,h^{-1} {\rm Mpc}$. This value reflects the typical void size obtained with the chosen void-finding algorithm and smoothing scale. We have explicitly verified that, for this void finder and smoothing parameter, the peak of the radius distribution exhibits little evolution with redshift. The bottom panel shows the normalised redshift distribution of the void catalogue as a histogram, overlaid with vertical dashed lines indicating the redshift bins used in our subsequent analysis.

\subsection{Galaxy colour selection}

As previously mentioned, the impact of intrinsic alignments strongly depends on galaxy type:
red galaxies tend to exhibit strong tidal alignment with the surrounding gravitational field, whereas blue galaxies show only weak or negligible alignment.
Consequently, it is common to divide galaxy catalogues into red and blue sub-samples. In this analysis, unlike the approach of \cite{Paviot}, who applied colour cuts based on absolute magnitudes (e.g., \( M_u - M_r > 1.32 \)), we adopt a different selection method.
Galaxies are classified directly using the \texttt{color\_kind} column provided in the Flagship simulations \citep{Flagship}, which allows us to separate red, blue, and green galaxies. This classification follows the methodology of \cite{Skibba2009} and \cite{carretero2015}, based on a Gaussian fit on the $g-r$ colour-magnitude diagram. 
\begin{figure}
    \centering
    \includegraphics[width=1\linewidth]{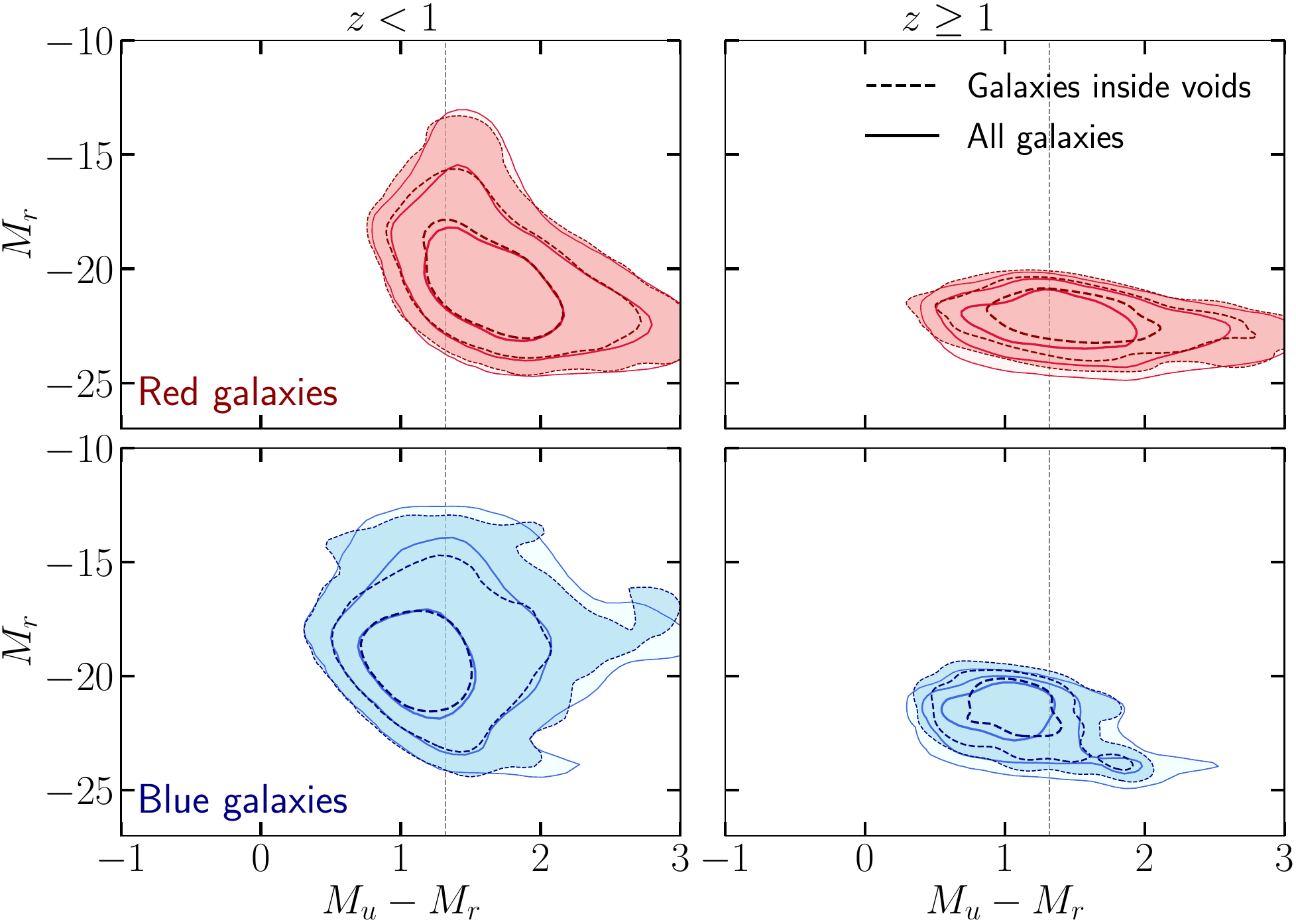}
    \caption{Colour-magnitude diagram for the full Flagship galaxy sample, with red and blue galaxies shown. Solid contours represent galaxies with redshift below $1$ (\textit{left panels}) and above $1$ (\textit{right panels}), while dashed contours correspond to galaxies residing inside voids. The dashed vertical line indicates the red-blue separation adopted by \cite{Paviot}.}
    \label{fig:galcolor}
\end{figure}

In Fig.~\ref{fig:galcolor}, we show the colour-magnitude diagram of galaxies classified using the \texttt{color\_kind} method, compared to the classification used in \cite{Paviot}, for redshifts below 
$1$ (left panels) and above $1$ (right panels). Solid contours represent the full distribution, while dashed contours show the diagram for galaxies inside voids (at distances from void centres smaller than the void radius $ R_\mathrm{V}$). The absolute rest-frame magnitudes $M_u$ and $M_r$ correspond to the CFHT {\it u} band and the Subaru {\it r} band, following the convention in \cite{Paviot}. The grey vertical dashed line indicates the $M_u-M_r = 1.32$ cut used to differentiate red and blue galaxies. Some differences in classification exist between the two methods, so some discrepancies between the analyses are expected. 
Nevertheless, the distribution suggests that the galaxy population inside voids is broadly similar to the overall population, allowing a meaningful comparison between the two signals.

\section{Method and concepts}\label{sec:methodandconcept}
\subsection{Gravitational lensing}
Gravitational lensing by cosmic voids has recently gained attention as a promising probe for future cosmological \citep{bonici2023,xiong2025}. In contrast to the lensing signals around massive clusters, the signal induced by voids is subtle due to their underdense nature. Void lensing is particularly sensitive to both cosmic geometry and the growth of structure, and it provides a unique window into modified gravity theories. In such theories, screening mechanisms suppress deviations from general relativity in high-density regions, while in underdense regions like voids, these effects are less suppressed and thus more readily observable.

During their initial observational efforts, several studies have reported statistically significant detections of the weak lensing signal around cosmic voids using galaxy survey data (e.g., \citealp{sanchez, fang, clampitt}), suggesting the feasibility of such measurements with current and upcoming data sets.

The observable quantity in void lensing is the excess surface mass density $\Delta \Sigma(r_\mathrm{p})$, which can be expressed in terms of the tangential shear profile $\gamma_\mathrm{t}(r_\mathrm{p})$ as
\begin{equation}
    \Delta\Sigma(r_\mathrm{p}) = \Sigma_{\mathrm{crit}} \, \gamma_\mathrm{t}(r_\mathrm{p})\, ,
\end{equation}
where $r_{\rm p}$ is the transverse component of the void-galaxy separation vector and $\Sigma_{\mathrm{crit}}$ is the critical surface mass density, defined as
\begin{equation}
    \Sigma_{\mathrm{crit}} = \frac{c^2}{4\pi \, G_\mathrm{N}} \, \frac{D_\mathrm{A}(z_\sfont{S})}{D_\mathrm{A}(z_\sfont{L}) \, D_\mathrm{A}(z_{\sfont{L}},z_{\sfont{S}}) \, (1+z_\sfont{L})^2}\, ,
\end{equation}
$G_\mathrm{N}$ being the Newton constant and with $D_\mathrm{A}(z_\sfont{S})$, $D_\mathrm{A}(z_\sfont{L})$, and $D_\mathrm{A}(z_{\sfont{L}},z_{\sfont{S}})$ denoting the angular diameter distances to the source with redshift $z_\sfont{S}$, to the lens with redshift $z_\sfont{L}$ (here the void), and between the lens and the source, respectively. The factor $(1+z_\sfont{L})^2$ accounts for the use of comoving coordinates in the analysis. Moreover, in the context of tomographic analyses, where source galaxies are divided into redshift bins, it becomes essential to account for the full redshift distribution of source galaxies within each bin. Rather than assuming a fixed source redshift, we compute an effective inverse critical surface density by integrating over the redshift probability distribution $n_{{\rm g},i}(z_\sfont{S})$ of source galaxies in redshift bin $i$,
\begin{equation}
    \Sigma^{-1}_{\mathrm{c, eff}}(z_{\sfont{L}}, i) = \int_{z_{\sfont{L}}}^{\infty} n_{{\rm g},i}(z_{\sfont{S}}) \, \Sigma^{-1}_{\mathrm{crit}}\paren{z_{\sfont{L}}, z_{\sfont{S}}} \, \diff z_{\sfont{S}}\, .
\end{equation}
This formulation ensures that the measured lensing signal properly accounts for the weighting of all source galaxies contributing to the shear measurement, allowing for a more accurate inference of the underlying matter distribution around voids.
For cosmic voids, wich are extended objects, the profile of $\Delta\Sigma(r_\mathrm{p})$ depends on the void radius. In this analysis, we select voids with radii in the range $10 < R_\mathrm{V}\,/(h^{-1} {\rm Mpc}) < 15$ to optimize statistical power (see Fig. \ref{fig:voidsraddist}). 

\subsection{Intrinsic alignment}
Intrinsic alignments refer to the physical correlations between galaxy shapes and the surrounding large-scale gravitational potential, arising from the influence of tidal gravitational fields on galaxy formation and evolution. 

Weak lensing aims to measure the coherent alignment of galaxy shapes induced by the gravitational deflection of light along the line-of-sight. Intrinsic alignments that are relevant for cosmological studies bias measurements of the weak lensing signal on galaxies.
Specifically, the observed shear $\gamma_{+/\times,i}^{\rm obs}$ of a given galaxy $i$  can be expressed as the sum of several contributions
\begin{equation}\label{eq:gammatot}
    \gamma^{\, \rm obs}_{+/\times,i}=\gamma_{+/\times,i}^{\, \rm I}+\gamma_{+/\times,i}^{\, \rm G} \, ,
\end{equation}
where $\gamma^{\, \rm I/G}_{+/\times}$ represents the alignment effects caused by intrinsic alignment (${\rm I}$) or cosmic shear (${\rm G}$). The correlation between the void positions $\delta_{\rm V}$ and galaxy alignments can be expressed as

\begin{equation}
    \langle \delta_{\rm V} \, \gamma^{\, \rm obs}_{+/\times} \rangle=\langle \delta_{\rm V} \, \gamma^{\, \rm G}_{+/\times} \rangle+\langle\delta_{\rm V} \, \gamma^{\, \rm I}_{+/\times} \rangle\, ,
\end{equation}
which typically shows that the observed shear signal can be contaminated by the intrinsic orientation of galaxies. In this analysis, we focus on modelling and measuring in the simulation the second term of the equation $\langle \delta_{\rm V}\gamma^{\, \rm I}_{+/\times} \rangle$.

Several theoretical models have been proposed to describe galaxies intrinsic alignments. The Linear Alignment (LA) model \citep{catelan2001} assumes a direct linear relation between galaxy ellipticity and the local gravitational tidal field. An extension of this framework, known as the nonlinear alignment (NLA) model \citep{hirata2004, bridle2007}, incorporates nonlinear corrections to better match observations on smaller scales. More recently, the tidal alignment and tidal torquing (TATT) model \citep{blazek2019} provides a more flexible parameterization, capturing both linear and quadratic contributions to galaxy alignments.

Cosmic voids, defined as the most underdense regions in the large-scale structure, offer a unique environment for studying galaxy intrinsic alignments. Galaxies located near void edges are expected to exhibit alignment patterns that trace the surrounding tidal field. In these low-density regions, where nonlinear structure formation effects are minimal \citep{Schuster2023,hamaus2014b}, galaxy shapes are predicted to respond approximately linearly to the tidal field, making the linear alignment model particularly appropriate. This expectation is supported by \citet{pollina2017}, who showed that the galaxy distribution around voids closely follows the predictions of linear bias theory, in contrast to overdense environments, where nonlinear dynamics play a significant role.

In this study, we focus on the linear alignment (LA) model. On large scales, LA model captures the intrinsic alignment signal of red galaxies and has been shown to provide a good description down to scales of $\sim 10\,h^{-1}\mathrm{Mpc}$ \citep{Chisari2014}, while its nonlinear extension (NLA) is generally required on smaller scales where nonlinear structure growth becomes important \citep{singh2015}. Given the typical size of voids and the predominantly linear regime probed around cosmic voids, the LA framework provides a well-motivated baseline model for our analysis \citep{d'Assignies,pollina2017}. Moreover, it provides a foundation for more complex models and can be extended to include redshift and luminosity dependencies.
In the LA model, the intrinsic-alignment shear field is first expressed in terms of its Cartesian components, 
$\gamma^{\rm I}_{\rm 1,2}(\vec{k})$, defined with respect to a fixed coordinate system. In Fourier space, these components depend linearly on the gravitational potential with an amplitude parameter $A_{\rm I}$,
\begin{equation}\label{eq:gammapm1}
\begin{split}
\gamma^{\rm I}_{\rm 1}(\vec{k}) &= -\frac{A_{\rm I} \, C_1}{4\pi \, G_\mathrm{N}} \, (k_x^2 - k_y^2) \, \Psi(\vec{k}) \,,\\
\gamma^{\rm I}_{\rm 2}(\vec{k}) &= -\, 2\frac{A_{\rm I} \, C_1}{4\pi \, G_\mathrm{N}}  \, k_x \, k_y \, \Psi(\vec{k}) \,,
\end{split}
\end{equation}
where $C_1=5 \times 10^{-14}\,M_\odot^{-1} \,{\it h}^{-2}\,\text{\rm Mpc}^{3}$ is a historical constant coming from the IA measurement in the local Universe \citep{brown2002}, \( \Psi(\vec{k}) \) is the gravitational potential in Fourier space, with the three-dimensional wave number \( \vec{k} = (k_x, k_y,k_z) \), and \( k^2 = k_x^2 + k_y^2 +k_z^2\). The gravitational potential is related to the matter density contrast \( \delta_{\rm m}(\vec{k}) \) by\footnote{This relation is valid only for non-zero Fourier modes (\(k\neq0\)); the \(k=0\) mode is excluded due to the mass-sheet degeneracy.}
\begin{equation}
\Psi(\vec{k}) = -\frac{4\pi \, G_\mathrm{N}\, \bar{\rho}_{{\rm m}0}\, \delta_{\rm m}(\vec{k})}{D(z)\, k^2}\, , \label{eq:grav_pot}
\end{equation}
where \( \bar{\rho}_{{\rm m}0} \) is the present-day mean matter density, and \( D(z) \) is the linear growth factor at redshift \( z \). Substituting the expression for \( \Psi(\vec{k}) \) into Eq. \eqref{eq:gammapm1} we obtain the combined expression
\begin{equation}\label{eq:gammapm}
\begin{split}
\gamma^{\rm I}_{\rm 1}(\vec{k}) &= A_{\rm I} C_1 \frac{k_x^2 - k_y^2}{k^2} \, \frac{\bar{\rho}_{{\rm m}0}}{D(z)} \, \delta_{\rm m}(\vec{k})\,,\\
\gamma^{\rm I}_{\rm 2}(\vec{k}) &= A_{\rm I} C_1 \frac{2\,k_x\, k_y}{k^2} \, \frac{\bar{\rho}_{{\rm m}0}}{D(z)} \, \delta_{\rm m}(\vec{k})\,.
\end{split}
\end{equation}

We then want to express the 2-point correlation function $\xi^{\, \rm I}_{\rm V,1/2}(\vec{r})$ in real space between the void positions $\delta_{\rm V}(\vec{r})$ and the galaxy IA using an inverse Fourier transform as
\begin{align}
\xi^{\rm I}_{\rm V,1/2}(\vec{r})
&= \langle \delta_{\rm V}(\vec{r})\,\gamma^{\rm I}_{\rm 1/2}(\boldsymbol{r}) \rangle \\
&= \int \frac{\mathrm{d}^3 k}{(2\pi)^3}\;
\mathrm{e}^{\mathrm{i}\,\boldsymbol{k}\cdot\boldsymbol{r}}\;
\langle \delta_{\rm V}^{*}(\boldsymbol{k})\,\gamma^{\rm I}_{\rm 1/2}(\boldsymbol{k})\rangle.
\notag
\end{align}
By inserting Eq. \eqref{eq:gammapm} we then obtain
\begin{align}\label{eq:xipxim}
\begin{split}
\xi^{\, \rm I}_{\rm V,1}(\vec{r}) &=
A_{\rm I}\,C_1 \, \frac{\bar{\rho}_{{\rm m}0}}{D(z)}
\\
&\quad \times\int \frac{\diff ^3 k}{(2\pi)^3} \;
{\rm e}^{{\rm i} \vec{k} \cdot \vec{r}} \;
\frac{k_x^2 - k_y^2}{k^2} \;
\big\langle \delta_{\rm V}^{*}(\vec{k}) \, \delta_{\rm m}(\vec{k}) \big\rangle , \\[4pt]
\xi^{\, \rm I}_{\rm V,2}(\vec{r}) &=
A_{\rm I}\,C_1\,\frac{\bar{\rho}_{{\rm m}0}}{D(z)}
\\
&\quad \times\int \frac{\diff ^3 k}{(2\pi)^3} \;
{\rm e}^{{\rm i}\vec{k}\cdot\vec{r}} \;
\frac{2\,k_x\,k_y}{k^2} \;
\ave{\delta_{\rm V}^{*}(\vec{k})\,\delta_{\rm m}(\vec{k})}\, .
\end{split}
\end{align}
Using statistical homogeneity and isotropy, the cross-power spectrum is defined as
\begin{equation}
\ave{\delta_{\rm V}^{*}(\vec{k}) \, \delta_{\rm m}(\vec{k}')}
= (2\pi)^3 \, \delta_{\rm D}^{(3)}(\vec{k}-\vec{k}')\,P_{\rm Vm}(k),
\end{equation}
which, after integrating over $\vec{k}'$, replaces the ensemble average by $P_{\rm Vm}(k)$ in Eq.~\eqref{eq:xipxim}.

For convenience in both theory and observations, we decompose the separation vector \(\vec{r}\) into a transverse component $r_\mathrm{p}$ and a line-of-sight component $\Pi$. 
This parametrization naturally captures anisotropic correlations and simplifies 
integration in Fourier space. Using cylindrical Fourier coordinates $(k_\perp,\theta,k_z)$, where 
$\diff ^3k = k_\perp\,\diff k_\perp\,\diff \theta\,\diff k_z$, the geometric kernel becomes
\begin{equation}
\frac{k_x^2-k_y^2}{k^2} = \frac{k_\perp^2}{k^2} \, \cos{2\theta}\, ,
\end{equation}
and the transverse plane wave is
${\rm e}^{{\rm i}k_\perp r_\mathrm{p}\cos\theta}$. Using the identity
\begin{equation}
\int_{0}^{2\pi}{\rm e}^{{\rm i}k_\perp r_\mathrm{p}\cos\theta}\cos{2\theta}\,\diff \theta
= -\,2\pi\,J_2(k_\perp \, r_\mathrm{p})\, ,
\end{equation}
and noting that $P_{\rm Vm}(k)$ depends only on 
$k=\sqrt{k_\perp^2+k_z^2}$ by isotropy, the $\theta$-integration can be 
performed analytically. At this stage, the shear has been rotated into the tangential/cross basis relative to the projected separation vector $r_\mathrm{p}$, so the $+$ and $\times$ subscripts refer to tangential and cross components, respectively.This reduces the expression to :  
\begin{align}\label{eq:lastxieq}
\xi^{\, \rm I}_{\rm V+}&(r_\mathrm{p},\Pi) = 
-\,A_{\rm I}\,C_1\,\frac{\bar{\rho}_{{\rm m}0}}{\pi^2\,D(z)}
\int_{0}^{\infty}\!\diff k_\perp  \int_{0}^{\infty}\!\diff k_z\;\nonumber\\
&\quad \times
\frac{k_\perp^{3}}{k_\perp^2 + k_z^2}\,
P_{\rm Vm}\! \paren{\sqrt{k_\perp^2+k_z^2}}  J_2(k_\perp r_\mathrm{p})\,\cos{(k_z \,\Pi)}\,.
\end{align}

For the cross component, the kernel $2k_xk_y/k^2$ integrates to zero by parity, 
yielding $\xi_{\rm V\times}(r_\mathrm{p},\Pi)=0$ in the isotropic case.

\subsection{Projected correlation functions}\label{sec:model}

In observations, we measure projected quantities to mitigate systematic errors. 
By integrating the 3-dimensional correlation function $\xi^{\, \rm I}_{\rm V+}(r_\mathrm{p}, \Pi)$ 
along the line-of-sight direction, we obtain the projected correlation function
\begin{equation}\label{eq:wpint}
w^{\, \rm I}_{\rm V+}(r_\mathrm{p}) = \int_{-\Pi_\mathrm{max}}^{\Pi_\mathrm{max}} 
\xi^{\, \rm I}_{\rm V+}(r_\mathrm{p}, \Pi) \, \diff \Pi \, .
\end{equation}
This projection mitigates the impact of photometric redshift uncertainties and redshift-space distortions (RSD), both of which can misplace objects 
along the line-of-sight and bias 3-dimensional measurements. Since the IA signal is, by definition, concentrated near the lensed objects, we adopt $\Pi_{\mathrm{\min}/\mathrm{\max}}=\pm\, 100\,h^{-1} {\rm Mpc}$ to ensure that the signal around our voids is fully captured. In the following, we model the amplitude and shape of this projected IA signal based on the underlying void-matter cross-power spectrum, which can be decomposed into distinct contributions as outlined by  \cite{d'Assignies} as
\begin{enumerate}
    \item the intermediate-scale regime, which characterizes the matter distribution near the void centre, where galaxy shapes tend to align tangentially around voids, producing a negative signal.
    \item the large-scale regime, which describes the large-scale matter distribution around cosmic voids, where galaxy shapes tend to align radially, resulting in a positive signal.
\end{enumerate} 

\subsubsection{\textit{Intermediate-scale regime (ISR)}}

Inside the void radius, the void-matter power spectrum can be approximated by
$\Delta_{\rm V}(k) \equiv P_{\rm Vm}(k)$, which is given by the Fourier transform of the void-density profile. On these scales, replacing the power spectrum in Eq.~\eqref{eq:lastxieq} by $\Delta_{\rm V}(k)$ and integrating over $\Pi$ to obtain the projected function as in Eq.~\eqref{eq:wpint}, we find
\begin{align}\label{eq:1voiterm}
w^{\, \rm I}_{\rm V+}(r_\mathrm{p}) &= 
A_{\rm I}^{\, \rm V} \, \frac{C_1\,\rho_{\rm crit}\,\Omega_{\rm m}}{\pi^2}\nonumber\\
& 
\times
\int_0^{+\infty} \! \diff z \; \frac{W(z)}{D(z)} \nonumber \int_0^{+\infty} \! \diff k_{\perp} \int_0^{+\infty} \! \diff k_z \;
\\
& 
\times\frac{k_{\perp}^3}{k^2\,k_z} \;
\Delta_{\rm V}(k) \, \sin{(k_z \,\Pi_{\mathrm{max}})} \,
J_2(k_{\perp} \, r_\mathrm{p}) \,,
\end{align}
with $\rho_{\rm crit}\,\Omega_{\rm m}=\bar{\rho}_{{\rm m}0}$ and where the redshift integral accounts for all void-galaxy pairs within the chosen redshift range, weighted by $W(z)$, thereby computing $w^{\, \rm I}_{\rm V+}(r_\mathrm{p})$ as an average over that range. The void-galaxy window function is defined as
\begin{equation}
    W(z) \;=\; \frac{p_{\rm V}(z)\,p_{\rm g}(z)}{\chi^2 \, \diff \chi/\diff z}
    \brackets{\int_0^{+\infty} \! \frac{p_{\rm V}(z)\,p_{\rm g}(z)}{\chi^2\,\diff \chi/\diff z} \, \diff z} ^{-1} \, ,
\end{equation}
where $p_{\rm V}(z)$ and $p_{\rm g}(z)$ are the redshift distributions of voids and galaxies, respectively (see \citealp{Mandelbaum2011}) and $\chi$ represents the comoving distance.
Thus, modelling the IA signal from cosmic voids on these scales requires the internal void-matter distribution $\Delta_{\rm V}(k)$, which we infer directly from the lensing signal of our cosmic voids, as described in the next section.

\subsubsection{\textit{Large-scale regime (LSR)}}
In this regime, we correlate voids centres with the intrinsic alignment signal of galaxies around the voids. This term is based on the theoretical framework developed by \cite{hamaus2014}, where the void-galaxy position power spectrum is modeled analogously to the two terms used for dark matter halos.
In this perspective, on large scales, cosmic voids can be treated as biased tracers of the underlying matter field.  The  LSR term therefore depends on the large-scale linear void bias $b_{\rm V}(R_{\rm V})$, and the void-matter power spectrum can be expressed in terms of the linear-matter power spectrum $P^{\mathrm{lin}}_{{\rm mm}}(k)$ as
\begin{equation}
    P_{\rm Vm}(k) \;=\; b_{\rm V}(R_{\rm V}) \, P^{\mathrm{lin}}_{\rm mm}(k) \, .
\end{equation}
The projected correlation function at large scales then becomes
\begin{align}\label{eq:2voidterm}
w^{\, \rm I}_{\rm V+}(r_\mathrm{p}) &= \nonumber
A_{\rm I}^{\, \rm V} \, b_{\rm V}(R_\mathrm{V}) \, \frac{C_1\,\rho_{\rm crit}\,\Omega_{\rm m}}{\pi^2}
\\
&\times \int_0^{+\infty} \! \diff z \; \frac{W(z)}{D(z)} \nonumber \int_0^{+\infty} \! \diff k_{\perp}
\int_0^{+\infty} \! \diff k_z \;
\\
& \times\frac{k_{\perp}^3}{k^2\,k_z} \;
P^{\mathrm{lin}}_{\rm mm}(k,z) \,
 \sin{(k_z \Pi_{\mathrm{max}})} \,
J_2(k_{\perp} r_\mathrm{p}) \,.
\end{align}

We have verified that both galaxy populations, inside and outside the voids, have consistent colour-magnitude distributions (Fig. \ref{fig:galcolor}). Given that the IA calibration in FS2 depends mainly on these parameters besides the galaxy redshift, we adopt a single $A^{\, \rm V}_{\mathrm{I}}$ parameter for both the ISR and LSR terms in this analysis. However, it is important to note that such verification will be crucial when analyzing observational data, as several studies have shown that galaxy properties can differ in void environments (see \citealp{Habouzit}; \citealp{papini}).

\subsubsection{Model parameters}
By jointly analyzing measurements on both small and large scales, we can constrain different parameters within our modelling framework.  
On small scales, the amplitude of the IA signal, $A_{\rm I}$, is inferred from the ISR term. In this context $A_{\rm I}$ acts as a free bias parameter linking the local tidal field to the strength of the IA signal. In weak lensing cosmological analyses, $A_{\rm I}$ is typically treated as a nuisance parameter and marginalized over to prevent biases in cosmological parameter estimation.  

The potential redshift evolution of $A_{\rm I}$ has been considered in previous studies. While \citet{schmitz2018} suggested that such an evolution might occur and \citet{yao2020} reported mild evidence for it, more recent works have found no compelling observational support \citep{fortuna2021,Samuroff2023,navarro2026}. Current constraints remain insufficiently precise to definitively confirm or rule out redshift evolution. Following a common parameterization, we model this potential evolution as
\begin{equation}\label{eq:Az}
    A_{\rm I}(z) = A_z \paren{ \frac{1+z}{1+z_{\rm p}}}^{\eta_1} \, ,
\end{equation}
where $A_z$ is the amplitude at the pivot redshift $z_{\rm p}$, and $\eta_1$ quantifies the redshift dependence. We adopt $z_{\rm p} = 0.62$, consistent with \cite{Paviot}, to facilitate direct comparison with their results.
On large scales, combining the small-scale (ISR) and large-scale (LSR) measurements enables constraints on the large-scale linear bias of cosmic voids, since the ISR term constrains the amplitude parameter $A_{\rm I}^{\, \rm V}$, while the LSR term constrains the combination $A_{\rm I}^{\, \rm V}\, b_{\rm V}$ (see Eqs.~\ref{eq:1voiterm} and \ref{eq:2voidterm}), thereby allowing us to break the degeneracy and infer $b_{\rm V}$.

In this work, we focus exclusively on the possible redshift evolution of the IA amplitude and do not attempt to model or constrain any luminosity dependence. Such a dependence has been commonly explored; for instance, \cite{Paviot} examined luminosity trends in detail, but we leave a full luminosity-dependent analysis for future work. We note, however, that part of the observed redshift evolution may be driven by an underlying luminosity dependence, as higher-redshift samples tend to include intrinsically more luminous galaxies, which are known to exhibit stronger IA signals.

%--------------------------------------------------------------------

\section{Estimators and covariance}\label{sec:estimator}

\subsection{IA estimation}\label{par:IAest}

The projected correlation function quantifies the excess probability, relative to a random distribution, of finding pairs of tracers (here, void centres and galaxies) separated by a given projected comoving distance \( r_\mathrm{p} \). It is computed by integrating the full 3-dimensional 2-point correlation function, \( \xi(r_\mathrm{p}, \Pi) \), along the line-of-sight direction \(\Pi\), as given in Eq. \eqref{eq:wpint}. This projection removes the impact of redshift-space distortions and peculiar velocities, which mainly affect line-of-sight separations, thus yielding a cleaner estimate of the underlying real-space clustering.

In our analysis, the 3-dimensional correlation function is computed directly from the simulation using the Davis\textendash Peebles estimator~\citep{davispeebles}, as implemented in the publicly available \texttt{TreeCorr} package.\footnote{\href{https://rmjarvis.github.io/TreeCorr}{https://rmjarvis.github.io/TreeCorr/}} 
To measure the tangential orientation of galaxies around voids, such as the void-galaxy IA signal $w^{\, \rm I}_{\rm V+}(r_\mathrm{p})$, we use the tangential component of the galaxy's intrinsic ellipticity $\gamma^{\, \rm I}_+$ (see Eq. \ref{eq:gammatot}) defined with respect to the void centre. This quantity reflects whether galaxies preferentially align tangentially around the void or radially toward it. Averaging over all void-galaxy pairs yields the estimator
\begin{equation}\label{eq:wpestim}
w^{\, \rm I}_{\rm V+}(r_\mathrm{p})
= \frac{1}{N_{\rm pairs}}
\sum_{i=1}^{N_{\rm pairs}} \gamma^{\,\rm I}_{+,i}\, .
\end{equation}
A positive value of this statistic corresponds to radial alignment, whereas a negative value reflects tangential alignment.\footnote{In the simulation, we can separately access the ellipticity induced by gravitational shear [\,$\gamma_i^{\mathrm{G}}$\,] and the intrinsic component [\,$\gamma_i^{\mathrm{I}}$\,] for each galaxy (see Eq. \ref{eq:gammatot}).
} Here, \(N_{\rm pairs}\) denotes the total number of void-galaxy pairs. This approach extends the Davis\textendash Peebles framework from simple number counts to shape correlations without requiring a random catalogue correction, since survey geometry effects are negligible for such a sparse void population. In galaxy lensing (or void lensing), random catalogues are typically used to correct for survey geometry, depth variations, and photometric redshift uncertainties.
They are used to remove spurious shear signals induced by the survey mask and to estimate the boost factor, which accounts for misplacement effects arising from photometric redshift uncertainties.
In our case, the measurements are performed on simulated data with uniform geometry, no photometric redshift uncertainties, and a simple mask.
We restrict the integration over \(\Pi\) to the range
\begin{equation}
|\Pi| \leq 100\, h^{-1} {\rm Mpc} \, ,
\end{equation}
relative to the void centre. This choice ensures that we focus on scales where the IA signal is expected to be dominant, while contamination from weak gravitational lensing remains relatively small. Extending the integration beyond this range would introduce a larger fraction of uncorrelated pairs, diluting the measured IA signal and lowering the overall signal-to-noise ratio. The transverse separations \(r_\mathrm{p}\) are binned into 30 logarithmically spaced intervals covering
\begin{equation}
 1 \leq r_\mathrm{p} \, /(h^{-1} {\rm Mpc}) \leq 600 \, ,    
\end{equation} 
enabling us to probe both the small-scale clustering inside or near voids (ISR) and the large-scale correlations in their surrounding environment (LSR).

\subsection{Inferring the dark matter density profile of voids from the void–lensing signal}\label{sec:voidsprofrecon}
As noted earlier, a key ingredient in modelling the projected ISR term correlation function lies in the Fourier-space density profile $\Delta_{\rm V}(k)$. In configuration space, this quantity is defined as the spherically averaged matter density contrast around void centres,
\begin{equation}
\Delta_{\rm V}(r) \equiv \frac{\rho_{\rm m}(r) - \bar{\rho}_{\rm m}}{\bar{\rho}_{\rm m}} = \xi_{{\rm V m}}(r),
\end{equation}
where $\rho_{\rm m}(r)$ is the mean matter density at distance $r$ from the void centre, $\bar{\rho}_{\rm m}$ is the cosmic mean matter density, and $\xi_{{\rm V m}}(r)$ denotes the void–matter cross-correlation function.
In the following, we describe the methodology used to infer this profile.
\subsubsection{Theoretical framework}

In contrast to the approach we here adopt in \cite{d'Assignies}, where the void density profile $\Delta_{\rm V}(r)$ was estimated empirically, we here adopt a direct reconstruction method that leverages an independent observable: the weak gravitational lensing signal around cosmic voids. Specifically, we aim to infer the projected dark matter distribution from the measured excess surface mass density, $\Delta\Sigma$, following the formalism presented in \cite{baldauf}, \cite{Boschetti}, and \cite{dvornik2018}.

In this framework, the lensing signal can be related to the projected matter distribution $\Upsilon(r_\mathrm{p})$ via the annular differential surface density (ADSD), defined as
\begin{equation}\label{eq:ups1}
    \Upsilon_{\mathrm{lensing}}(r_\mathrm{p}, R_0) = \Delta\Sigma_{\rm t}(r_\mathrm{p}) - \frac{R_0^2}{r_\mathrm{p}^2} \, \Delta\Sigma_{\rm t}(R_0)\, ,
\end{equation}
where $r_\mathrm{p}$ is the projected comoving separation from the void centre, $R_0$ is an inner cutoff scale introduced to mitigate the impact of small-scale systematic effects, and $\Delta\Sigma_{\rm t}$ denotes the tangential excess surface mass density.

The corresponding quantity for the underlying dark matter distribution is given by
\begin{align}\label{eq:ups2}
\Upsilon_{\mathrm{DM}}(r_\mathrm{p}, R_0)
&= \frac{2}{r_\mathrm{p}^2} \brackets{\int_{R_0}^{r_\mathrm{p}} \mathrm{d}r'\, r' \, \Sigma(r')} \\
&\quad - \frac{1}{r_\mathrm{p}^2} \brackets{ r_\mathrm{p}^2 \, \Sigma(r_\mathrm{p}) - R_0^2 \, \Sigma(R_0)},
\notag
\end{align}
where $\Sigma(r_\mathrm{p})$ is the projected surface mass density of matter around voids.

From Eq.~\eqref{eq:ups2}, $\Sigma(r_\mathrm{p})$ can be recovered through differentiation,
\begin{equation}\label{eq:sigma}
    \frac{\diff}{\diff r_\mathrm{p}} \left[ \Upsilon(r_\mathrm{p}) \, r_\mathrm{p}^2 \right] \, \frac{1}{r_\mathrm{p}^2} = -\frac{\diff \Sigma(r_\mathrm{p})}{\diff r_\mathrm{p}}\, .
\end{equation}
This procedure therefore enables a direct reconstruction of the 2-dimensional dark matter profile as a function of the projected separation. It is worth noting that \cite{Boschetti} validated this relation for a specific void definition, which differs from the one used here. In Appendix \ref{ap:dm_dsig}, we demonstrate that the methodology is robust across different void finder techniques, supporting its application in the present analysis.

\subsubsection{Application to numerical simulation}
\begin{figure}
    \centering
    \includegraphics[width=1\linewidth]{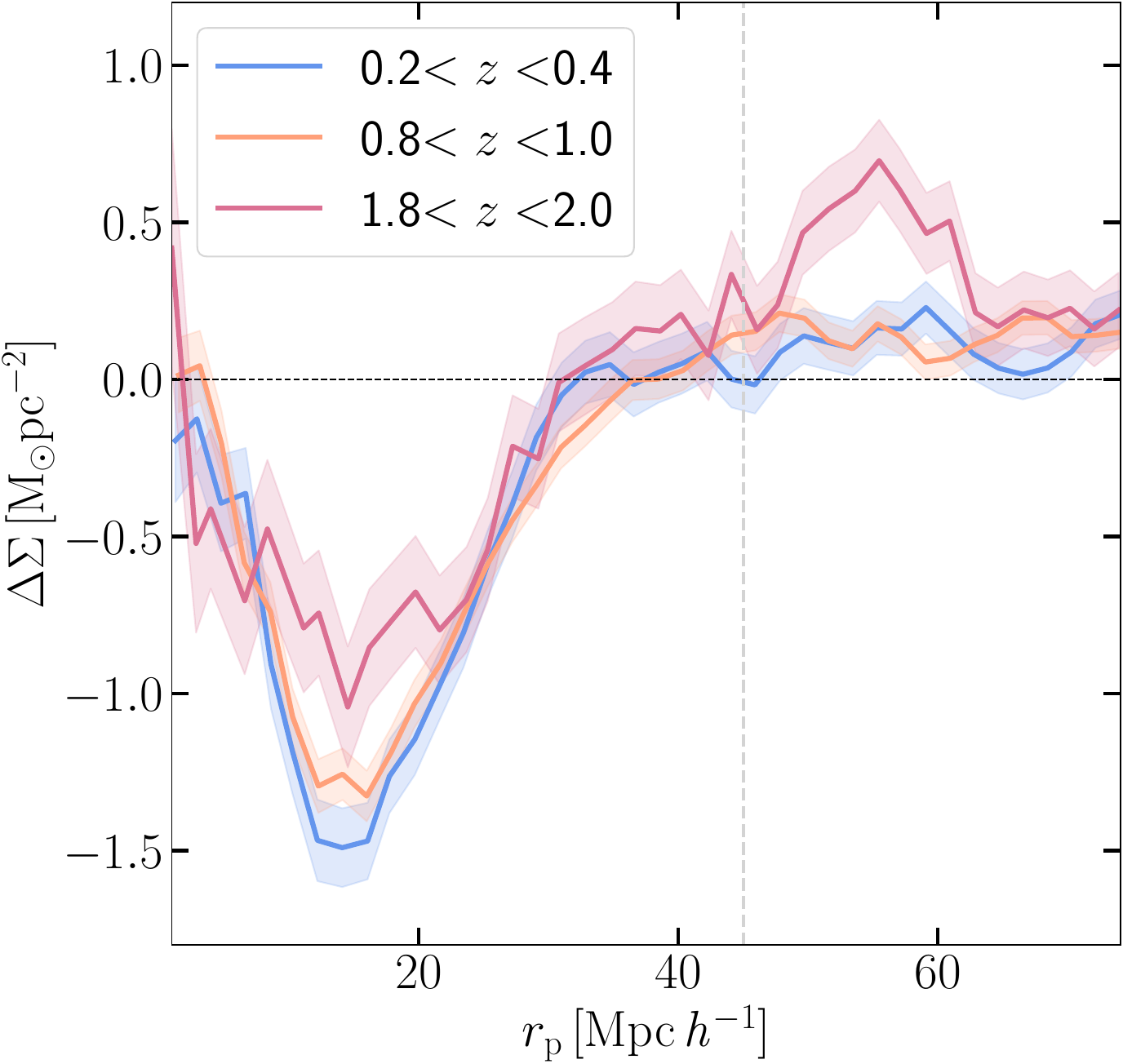}
    \caption{Lensing signal of galaxies for cosmic voids in the Flagship simulation, shown for three different redshift bins and for voids with radius in the range $10<R_\mathrm{V}\,/(h^{-1} {\rm Mpc})<15$. The grey dashed vertical line indicates the cutoff applied for the void-lensing profile reconstruction discussed in Sect. \ref{sec:voidsprofrecon}.}
    \label{fig:deltasig2}
\end{figure}

We measure the lensing signal of the 2D cosmic voids identified in our numerical simulation using the same estimator as before (Eq. \ref{eq:wpestim}) with the \texttt{TreeCorr} algorithm. The measured signal has been processed using background galaxies (with higher redshift than the considered voids) in the Flagship simulation up to redshift $z=2$. In Fig. \ref{fig:deltasig2}, we show the lensing signal induced by our cosmic voids for different redshift bins. A strong negative signal is observed within the void radius, consistent with theoretical expectations and previous observations. The figure also illustrates that, at higher redshifts, the lensing signal around cosmic voids decreases in amplitude and becomes noisier.

We connect the projected void-density profile, which corresponds to the projected mass density $\Sigma(r_\mathrm{p})$, obtained by inserting the lensing signal from Eq. \eqref{eq:ups1} and integrating via Eq. \eqref{eq:sigma} to the underlying 3D spherical profile $\xi(r)$ using the Abel transform~\citep{abel,bracewell}, a technique that has been previously applied in the study of cosmic voids (see \citealp{pisani2014,hawken2017,hamaus2020}). Under the assumption of spherical symmetry, the projected profile is related to the 3D profile through the forward Abel transform
\begin{equation}
\Sigma(r_\mathrm{p}) = 2 \int_{r_\mathrm{p}}^{\infty} \frac{r \, \xi(r)}{\sqrt{r^2 - r_\mathrm{p}^2}} \, \diff r\, ,
\end{equation}
and the 3D profile can be recovered by applying the inverse Abel transform
\begin{equation}
\xi(r) = -\frac{1}{\pi} \int_r^{\infty} \frac{\diff \Sigma}{\diff r_\mathrm{p}} \, \frac{\diff r_\mathrm{p}}{\sqrt{r_\mathrm{p}^2 - r^2}}\, .
\end{equation}

\begin{figure}
    \centering
    \includegraphics[width=0.96\linewidth]{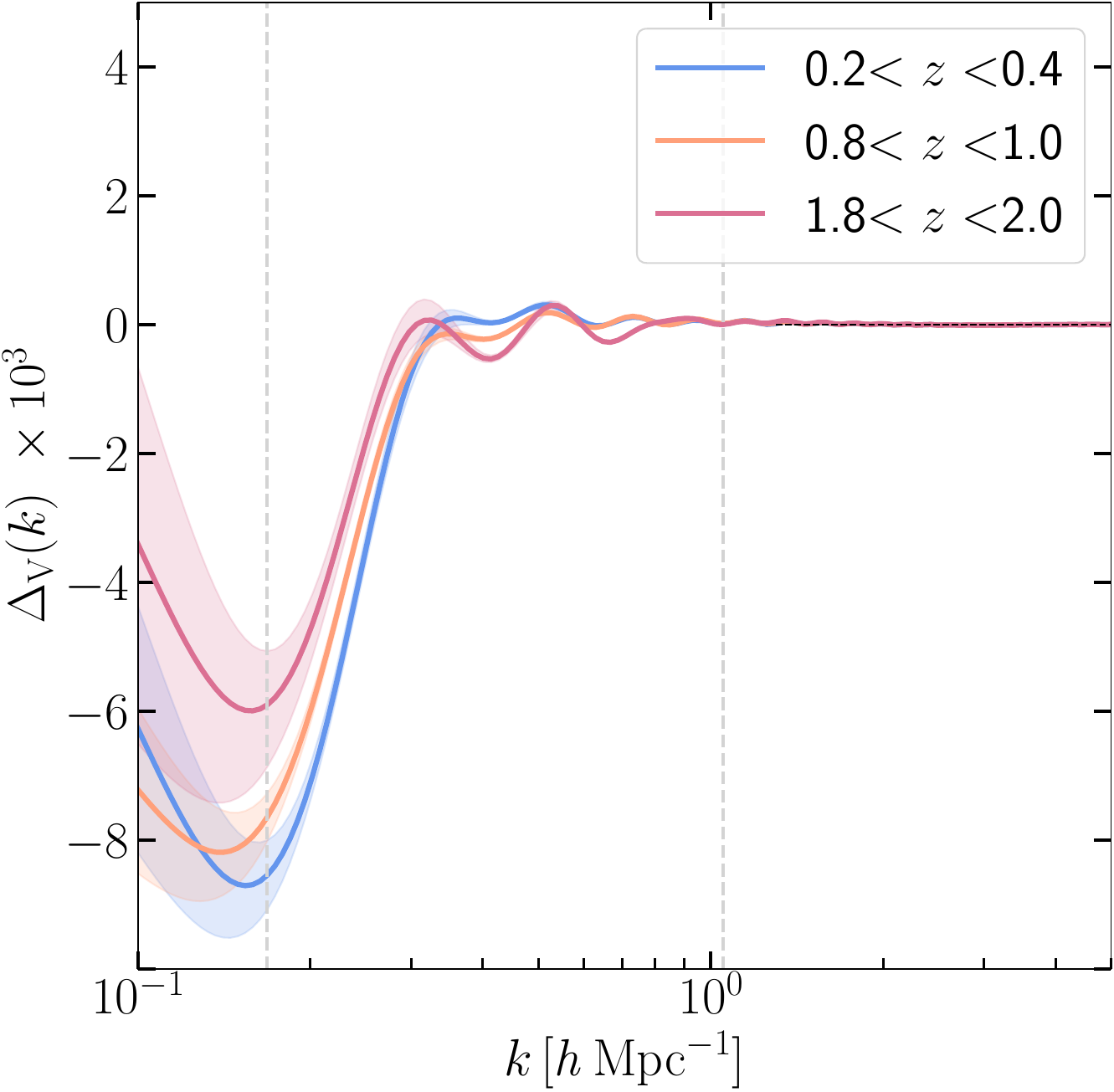}
    \caption{Reconstructed void density profile in Fourier space $\Delta_{\rm V}(k)$ for three different redshift bins. The grey dashed vertical line indicates the cutoff applied for the void-lensing profile reconstruction discussed in Sect. \ref{sec:voidsprofrecon}.}
    \label{fig:deltavk}
\end{figure}

The void profile in Fourier space is then inferred following the approach of \cite{chan2014},
\begin{equation}
    \Delta_{\rm V}(k)=\int_0^{+\infty}4\pi\, r^2 \, \diff r \frac{\sin{kr}}{kr}\,\Delta_{\rm V}(r)\, .
\end{equation}
We show in Fig.\,\ref{fig:deltavk} the reconstructed profile of our voids in Fourier space. We observe a shape of the Fourier space profile consistent with the results of \cite{chan2014}. As expected from the evolution of structure, we also find a redshift dependence of the signal amplitude, with deeper voids at lower redshift.

\subsection{Covariance and fitting methodology}\label{par:covmat}

The statistical uncertainties on $w^{\, \rm I}_{\rm V+}(r_\mathrm{p})$ are given by the square root of the diagonal terms of the covariance matrix $\mathbf{C}$, which is estimated via a jackknife resampling technique following the delete-$n_{\mathrm{patch}}$ scheme described in \cite{escoffier2016}. While the full covariance matrix, including off-diagonal terms, is used in parameter estimation, only the diagonal elements are shown as error bars in the figures for clarity. In this approach, the full survey or simulation volume is divided into \(N_s\) disjoint spatial patches of approximately equal area. A jackknife realization is obtained by omitting \(N_d\) of these patches, computing the statistic of interest on the remaining data, and repeating this process over all possible combinations of deleted patches. The covariance matrix is then computed as
\begin{equation}
\mathbf{C} = \frac{N_s - N_d}{N_d \, N_{\mathrm{jk}}} 
\sum_{i=1}^{N_{\mathrm{jk}}} 
\left( \mathbf{x}_i - \bar{\mathbf{x}} \right) 
\left( \mathbf{x}_i - \bar{\mathbf{x}} \right)^{\mathrm{T}}\, ,
\end{equation}
where \(\mathbf{x}_i\) represents the measurement of the \(i\)-th jackknife realization, \(\bar{\mathbf{x}}\) is the mean over all realizations, and \(N_{\mathrm{jk}}\) is the total number of jackknife samples. The prefactor \(\frac{N_s - N_d}{N_d \, N_{\mathrm{jk}}}\) ensures an unbiased covariance estimate for the delete-\(n\) method. In this work, we divide the volume into \(N_s = 12\) patches, omit \(N_d = 3\) patches per realization, yielding \(N_{\mathrm{jk}} = \binom{12}{3} = 220\) jackknife samples. This choice provides a balance between having sufficiently large sub-volumes to capture cosmic variance and generating enough realizations for a stable covariance estimate.

Moreover, in the analytical expression for the ISR term , the dark-matter void profile \(\Delta_{\rm V}(k)\) is inferred from measurements of the void weak lensing signal \(\Delta\Sigma\) (see Sect. \ref{sec:voidsprofrecon}). To account for uncertainties in this inference, we propagate the errors from the weak lensing measurement into a corresponding reconstruction uncertainty.
Specifically, we compute three versions of \(\Delta_{\rm V}(k)\) corresponding to \(\Delta\Sigma\), \(\Delta\Sigma + \sigma(\Delta\Sigma)\), and \(\Delta\Sigma - \sigma(\Delta\Sigma)\), where \(\sigma(\Delta\Sigma)\) is the measurement uncertainty. The theoretical uncertainty is then taken as half the range spanned by these three \(\Delta_{\rm V}(k)\) values. The square of these theoretical uncertainties are added to the diagonal elements of the covariance matrix used in the fitting procedure.

\section{Results}\label{sec:results}

\subsection{IA parameter}

To maximize the statistical significance of the IA measurement, we select voids with radii in the range \(10 < R_\mathrm{V} \,/(h^{-1} {\rm Mpc}) < 15\), corresponding to the peak of the void radius distribution (see Fig. \ref{fig:voidsraddist}), yielding a total of 44\,450 voids. These voids are then divided into redshift bins from $z = 0.2$ to $z = 2.0$ with widths of $\Delta z = 0.2$. Table~\ref{tab:redshiftbins} lists the number of voids in each bin. The variation in void counts across redshift bins reflects both the changing comoving volume and the non-monotonic redshift dependence of the tracer number density. We then compute the correlation between the voids described in Sect.~\ref{sec:void_cat}, selected by the radius range above, and the IA signal of galaxies located within a line-of-sight distance of $\pm 100\,h^{-1} {\rm Mpc}$ from the void centres. This measurement probes how the matter distribution in and around cosmic voids influences the orientations of surrounding galaxies.

The measurement is performed independently for the red and blue galaxy samples. The resulting signals are then modeled using the analytical two-term expression given in Eqs.~\eqref{eq:1voiterm} and \eqref{eq:2voidterm}. The fits are carried out over the following projected separation ranges:

\bi
    \item ISR term \textbf{(inside voids):} $6\, h^{-1} {\rm Mpc} < r_\mathrm{p} < 2.5\, R_\mathrm{V}$;
    \item LSR term \textbf{(around voids)}: $4\, R_\mathrm{V} < r_\mathrm{p} < 500\, h^{-1} {\rm Mpc}$.
\ei

\noindent
The ISR term range focuses on scales inside the void where the lensing profile remains sufficiently noise-free to enable a robust reconstruction of the void-density profile. We also note that “Inside voids” refers to the region within the void aperture extending beyond $R_\mathrm{V}$, thus including both the underdense core (up to $R_\mathrm{V}$) and the surrounding compensation wall.) We restrict the range to $r_\mathrm{p}$ greater than $6\,h^{-1} {\rm Mpc}$ to exclude the innermost void regions, where the scarcity of galaxies leads to noise-dominated measurements. As the IA signal is consistent with zero on these scales, including them would not tighten parameter constraints and could instead degrade the quality of the fit.
The LSR term is evaluated starting from \(4\,R_\mathrm{V}\) to ensure the measurement lies firmly within the large-scale linear regime, where the void bias can be modeled reliably. The two regimes intentionally do not overlap: the void profile reconstruction is truncated at $2.5\,R_{\rm V}$ because the lensing signal becomes too noisy at larger scales and could affect the reconstruction, while the LSR term is initiated at $4\,R_{\rm V}$ to ensure that the analysis remains within the linear bias regime.
For each bin, we infer both the IA parameter \(A_{\rm I}\), which quantifies the strength of alignment between galaxy shapes and the large-scale structure, and the void-bias parameter \(b_{\rm V}(z)\), which describes the clustering bias of voids relative to the matter distribution. Together, these parameters shed light on how void environments influence galaxy alignments across cosmic time.

\begin{figure*}[t]
\centering
\includegraphics[width=\textwidth]{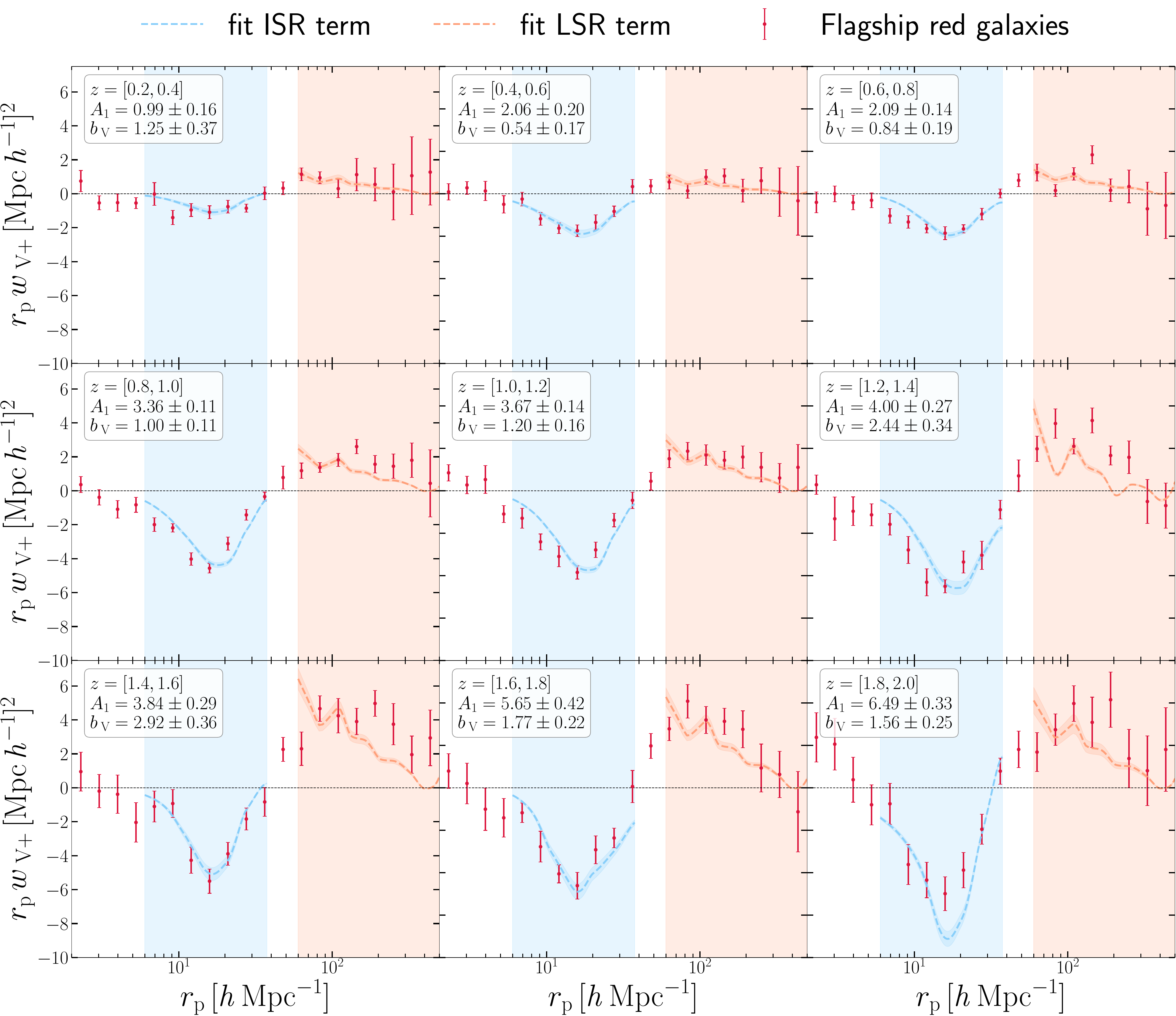}
\caption{Intrinsic alignment signal of the red-galaxy sample around void centres. Each panel represents a redshift bin. The blue and red shaded regions indicate the ranges used to fit the ISR and LSR terms, respectively. The dashed lines show the best-fit theoretical model, and the error bars are our measurements in the Flagship simulation.}
\label{fig:redshiftevol}
\end{figure*}

\begin{figure*}[htbp]
\includegraphics[width=\textwidth]{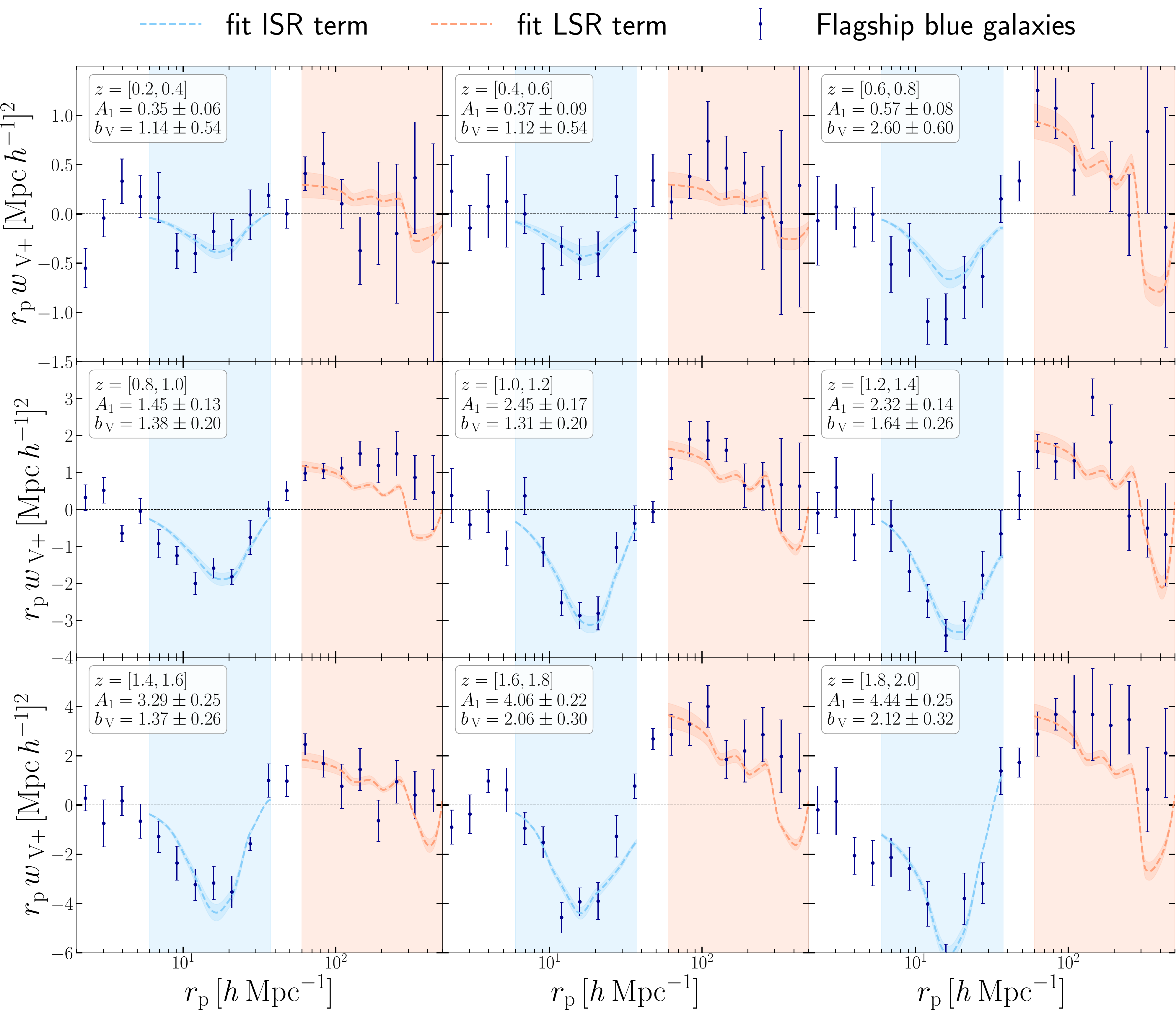}
 \caption{Same as Fig.\ref{fig:redshiftevol} for blue galaxies.}   

    \label{fig:redshiftevolblue}
\end{figure*}

Figures~\ref{fig:redshiftevol} and \ref{fig:redshiftevolblue} show the measured correlation functions, \( r_\mathrm{p} \, w^{\, \rm I}_{{\rm V+}}(r_\mathrm{p}) \), from the \textsc{Flagship} simulations, along with the best-fit ISR (blue) and LSR (red) contributions for voids in different redshift bins. As expected, a pronounced negative signal is observed within void interiors, indicating that galaxies tend to orient their major axes tangentially to the void centre, consistent with the stretching of structures along underdense regions. This transitions to a weaker positive signal in the outskirts, reflecting a preferential radial alignment relative to the void centre. This behaviour arises from the interplay between the compensating walls of voids and the surrounding large-scale tidal field. These results are in agreement with previous predictions and measurements, including the $95.45\%$ detection reported in \cite{d'Assignies} and the results of \cite{varela2012}. Notably, the amplitude of the ISR term increases with redshift for both galaxy samples, suggesting that alignments within void interiors are stronger at earlier cosmic times, likely due to the less-evolved dynamical state of structures at high redshift. 

The error bars represent the square root of the diagonal elements of the covariance matrix, estimated using the jackknife method described in Sect.~\ref{par:covmat}. We then fit the analytical model (Eqs.~\ref{eq:1voiterm} and~\ref{eq:2voidterm}) to the measurements in the ISR and LSR for each redshift bin using the \texttt{iminuit} algorithm.\footnote{\url{https://github.com/scikit-hep/iminuit}} The resulting best-fit values of \(A_{\rm I}\) and their uncertainties are reported in Table~\ref{tab:redshiftbins} for red and blue galaxies.

\begin{table*}[t]
\centering
\caption{Number of voids within radius $10 < R_{\rm V}\,/(h^{-1} {\rm Mpc}) < 15$ identified using the 2D void finder in the Flagship simulation for the different redshift bins. For each of these redshift bins, the best-fit values of the IA parameter $A_{\rm I}$ and the void bias $b_{\rm V}(z)$ obtained from the ISR and LSR terms are shown, separately for red and blue galaxies.}
\label{tab:redshiftbins}

\renewcommand{\arraystretch}{1.5} 
\resizebox{\textwidth}{!}{%
\begin{tabular}{lccccccccc}
\hline\hline
\textbf{Void redshift bin} 
    & $[0.2,0.4]$ & $[0.4,0.6]$ & $[0.6,0.8]$ & $[0.8,1.0]$ 
    & $[1.0,1.2]$ & $[1.2,1.4]$ & $[1.4,1.6]$ & $[1.6,1.8]$ & $[1.8,2.0]$ \\
\hline % ligne sous l’en-tête
\textbf{Number of voids} & 2079 & 3245 & 4021 & 6115 & 4742 & 4367 & 5662 & 7966 & 6253 \\
\hline % ligne en bas (optionnel)    
& & & & \textbf{Red galaxies} & & & & & \\
\hline
$A_{\rm I}^{\, \rm (red)} $ 
    & $1.02 \pm 0.13$ & $1.90 \pm 0.04$ & $2.46 \pm 0.08$ & $3.40 \pm 0.10$
    & $3.72 \pm 0.11$ & $3.31 \pm 0.22$ & $3.47 \pm 0.13$ & $5.42 \pm 0.25$ & $6.49 \pm 0.33$ \\
$b_{\rm V}(z)$ 
    & $1.22 \pm 0.34$ & $0.59 \pm 0.18$ & $0.71 \pm 0.15$ & $0.99 \pm 0.11$
    & $1.19 \pm 0.15$ & $2.94 \pm 0.41$ & $3.22 \pm 0.34$ & $1.84 \pm 0.20$ & $1.56 \pm 0.25$ \\
\hline    
& & & & \textbf{Blue galaxies} & & & & & \\
\hline
$A_{\rm I}^{\,\rm (blue)}$ 
    & $0.35 \pm 0.06$ & $0.37 \pm 0.09$ & $0.57 \pm 0.08$ & $1.45 \pm 0.13$
    & $2.45 \pm 0.17$ & $2.32 \pm 0.14$ & $3.29 \pm 0.25$ & $4.06 \pm 0.22$ & $4.44 \pm 0.25$ \\
$b_{\rm V}(z)$ 
    & $0.80 \pm 0.45$ & $1.03 \pm 0.52$ & $2.62 \pm 0.59$ & $1.37 \pm 0.19$
    & $1.14 \pm 0.17$ & $1.59 \pm 0.25$ & $1.31 \pm 0.22$ & $2.24 \pm 0.32$ & $2.28 \pm 0.32$ \\
\hline
\end{tabular}%
}

\end{table*}

\begin{figure}
    \centering
    \includegraphics[width=1\linewidth]{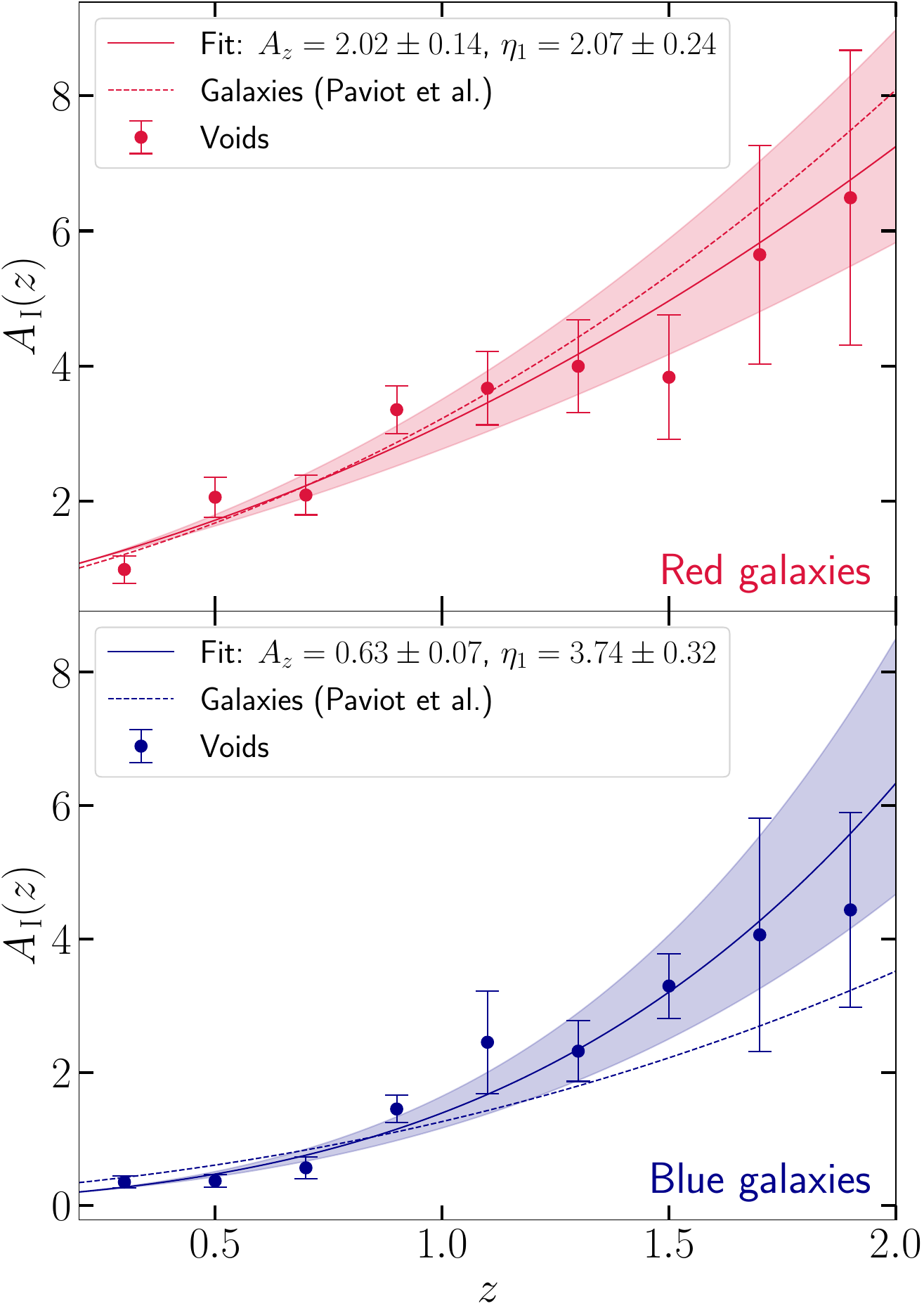}

    \caption{Redshift evolution of the amplitude of the IA signal measured around cosmic voids for red (\textit{top panel}) and blue (\textit{bottom panel}) galaxies in the Flagship simulation.}
    \label{fig:Az}
\end{figure}

Figure~\ref{fig:Az} shows the measured redshift evolution of the IA amplitude, \( A_{\rm I}(z) \), for red (top panel) and blue (bottom panel) galaxies, together with the best-fitting power-law model from Eq.~\eqref{eq:Az}. For red galaxies, the best-fit parameters are \( A_z = 2.02 \pm 0.11 \) and \( \eta_1 = 1.86 \pm 0.18 \), with a reduced $\chi^2$ of \( \chi^2/\mathrm{dof} = 1.93 \). For blue galaxies, we find \( A_z = 0.63 \pm 0.07 \) and \( \eta_1 = 3.74 \pm 0.32 \), with \( \chi^2/\mathrm{dof} = 1.01 \). As expected, the IA amplitude is significantly higher for red galaxies than for blue galaxies.\

For comparison, Fig.~\ref{fig:Az} also includes the corresponding measurements from \cite{Paviot}, who derived $A_{\rm I}(z)$ from the IA signal around galaxies rather than around voids. We find that the redshift evolution of the IA amplitude for both red and blue galaxies inferred from cosmic voids is in good agreement with the galaxy-based measurements, despite differences in the galaxy colour-separation criteria adopted in the two analyses. This consistency is expected when similar galaxy populations are considered.

\subsection{Bias evolution}
Unlike galaxies and clusters, which trace overdense regions, cosmic voids probe the low-density environments of the Universe, offering unique insights into cosmic expansion and gravity (\citealp{Moresco2022} and references therein). On large scales, corresponding to the LSR term, the void bias quantifies how voids trace the underlying matter distribution, indicating whether voids preferentially reside in regions with specific density fluctuations.

In the literature, cosmic voids can be broadly classified into two categories (\citealp{sheth2004}):  
\bi
    \item Void-in-voids: typically anti-biased voids that evolve within larger underdense regions;  
    \item Void-in-clouds: positively-biased voids that form in underdense regions embedded within overall overdense environments where galaxies cluster.
\ei
Although it is challenging to cleanly separate a void catalogue into these two populations, small, deep voids correspond to void-in-clouds structures, whereas larger, shallower voids are typically classified as void-in-voids (see \citealp{sheth2004}).
Studies such as \cite{hamaus2014} and \cite{Schuster2019} have shown that void bias becomes increasingly negative with larger void radii, while smaller voids tend to exhibit positive bias.

Understanding void bias is essential for cosmology because void properties are sensitive to the Universe’s expansion history and the nature of dark energy. Analogous to galaxy bias, void bias encapsulates how voids trace the underlying matter field, providing a complementary perspective on the interplay between matter and gravity on large scales. Moreover, void bias enters directly into theoretical models of void observables, including the void–matter correlation function and void–galaxy cross-correlations, enabling precise comparisons between observations and cosmological predictions \citep{hamaus2014,chan2014,Schuster2019}.

For voids with effective radii in the range $10 < R_\mathrm{V} \, /(h^{-1} {\rm Mpc}) < 15$, the bias is expected to evolve significantly with redshift. These smaller voids are most likely of the void-in-clouds type, and therefore tend to correlate positively with surrounding large-scale structures, such as galaxy clusters and haloes, which grow more prominent over time. At higher redshifts (\(z \gtrsim 1\)), collapsed structures are less abundant and less clustered, leading to a higher void bias. As the Universe evolves toward lower redshifts, increased clustering of these structures reduces the bias of small voids, since the enhanced growth of surrounding overdensities can shrink or even erase small, shallow voids. Consequently, the surviving small voids are typically located in denser environments, making them less anti-biased relative to the matter distribution.

Consequently, a decreasing trend of void bias with decreasing redshift is expected for this size range. 

\begin{figure}
    \centering
    \includegraphics[width=1\linewidth]{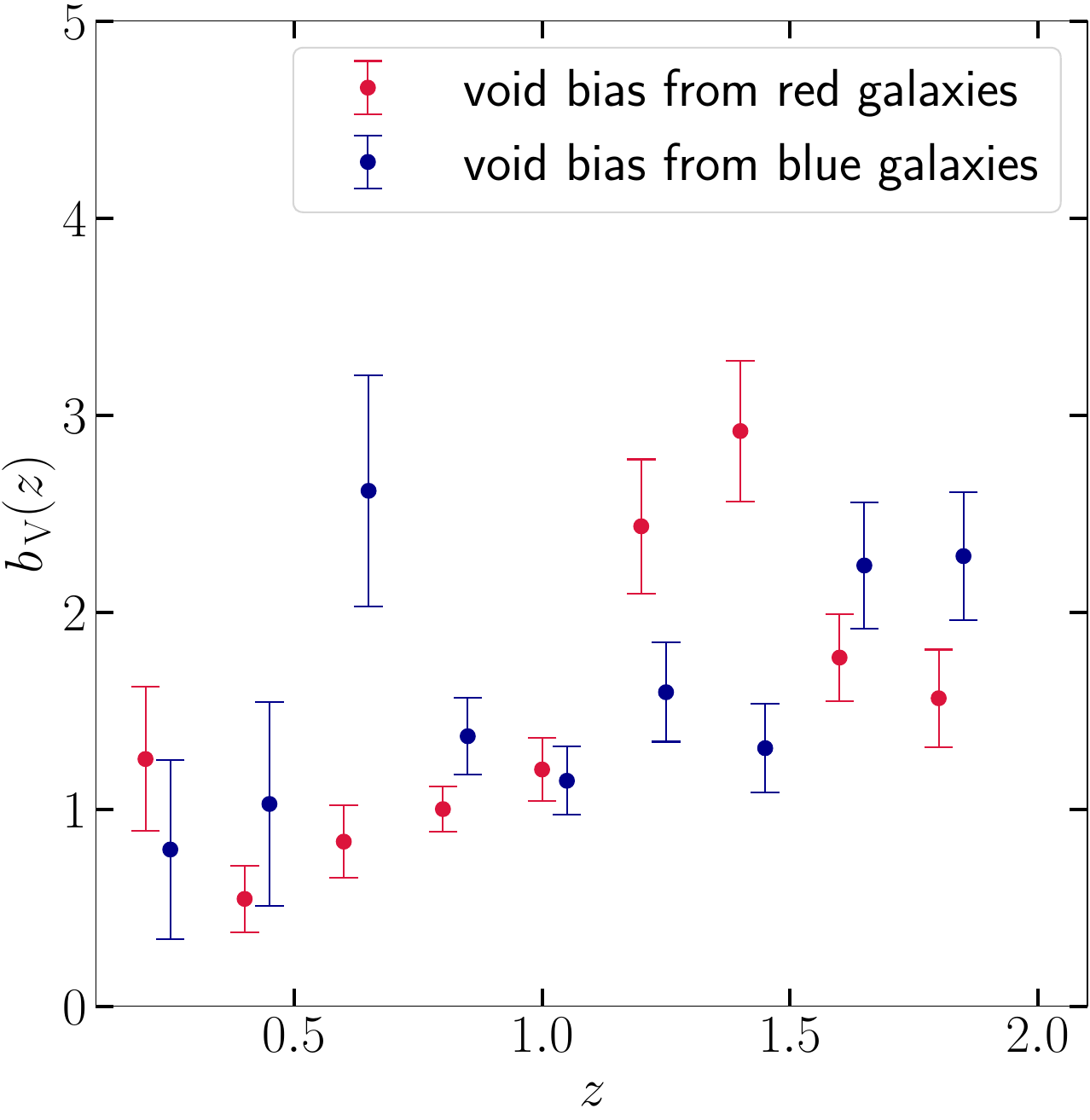}
    \caption{Redshift evolution of the linear-void bias $b_{\rm V}(z)$ measured from the void-IA signal for both blue and red galaxies in the Flagship simulation.}
    \label{fig:biasevol}
\end{figure}

Using the measurements presented here, we estimate the redshift evolution of void bias from both the ISR and LSR terms. Figure~\ref{fig:biasevol} shows this evolution for the IA signals measured in red and blue galaxy samples. We find good agreement between the two measurements, and, as expected, observe that void bias increases with redshift. However, we should emphasise that, given the use of 2D voids, the interpretation of these trends should remain qualitative, as this framework does not allow for a fully robust separation of different void sub-populations, and we leave a more detailed investigation of this aspect to future work.  

\section{Conclusions}\label{sec:conclusion}
The \textit{Euclid} mission will deliver precise measurements of the positions and shapes of billions of galaxies across the largest observable scales. This unprecedented data set will offer a unique opportunity to probe the fundamental nature of dark matter and dark energy. While galaxy-galaxy correlations will serve as a primary cosmological probe, cosmic voids present a powerful and complementary avenue, particularly well suited for testing alternative cosmological scenarios ( \citealp{bonici2023,contarini2022,Radinovic2023,hamaus2022}).

When using cosmic voids as tracers of the matter field, especially through void lensing, it is crucial to identify and control systematic effects that could bias the analysis. A key systematic in weak gravitational lensing measurements is the intrinsic alignment of galaxies, which arises from the tendency of galaxy shapes to align with the tidal fields in their local environments. If unaccounted for, this effect can mimic or contaminate the lensing signal. To date, the impact of intrinsic alignments on void–lensing measurements has not been extensively explored, and this work provides a detailed examination of this effect in the context of cosmic voids.

The aim of this work, in the context of the upcoming \Euclid observations, is to prepare and optimise the detection and characterization of the IA signal of red and blue galaxies in the environments of cosmic voids, as well as to study its redshift evolution. We use the Flagship cosmological simulation to construct a 2D void catalogue using a void-finding algorithm applied to the projected galaxy density field. The IA signal is measured separately within void interiors (the ISR term) and in their surrounding regions (the LSR term), for both red (elliptical) and blue (spiral) galaxies. Simultaneously, we implement a theoretical model for this signal, using the void-lensing correlation to directly infer the matter density profile inside voids. This approach enables a self-consistent probe of the matter distribution in and around voids without introducing additional free parameters. By relying directly on the lensing signal, we capture the key features of the matter field that drive the IA signal, ensuring the model remains closely tied to observations. Owing to the high-precision weak lensing measurements that \Euclid will provide, this framework can be applied over large cosmological volumes, enabling a robust reconstruction of the void-density profile and offering a direct observational window into the structure of underdense regions in the Universe.

In this analysis, we focus on voids with effective radii in the range $10 < R_\mathrm{V} \,/(h^{-1} {\rm Mpc}) < 15$, corresponding to the peak of the void-radius distribution in our catalogue. We then perform a study to quantify the redshift dependence of the IA amplitude for each galaxy population. Our measurements reveal a redshift evolution consistent with a power-law behaviour, in agreement with the findings of the companion study of \cite{Paviot}.

Moreover, the two-component theoretical model (ISR and LSR terms) enables an independent estimation of the bias of cosmic voids, measured separately for red and blue galaxies.  For the void size considered, we find a positive void bias that increases with redshift, in line with theoretical expectations from the evolution of large-scale structure. These findings highlight both the potential of void-galaxy IA measurements as a cosmological probe and the importance of accounting for IA as a systematic in future \textit{Euclid} weak lensing analyses similar to what has been found in galaxy-galaxy studies (\citealp{blot2025}, \citealp{navarro2026b}). Additionally, our results are consistent with previous studies \citep{varela2012,d'Assignies}, which report similar overall trends in galaxy intrinsic alignments.

\begin{acknowledgements}
This work was supported by CNES, focused on the \Euclid space mission.

\AckEC
\AckCosmoHub
 
\end{acknowledgements}

\bibliography{bib}

\begin{appendix} \label{app:lensingtoDM}
\section{Validity of the transformation between lensing and density profile}\label{ap:dm_dsig}
In our analysis, and particularly for the modelling of the IA signal around cosmic void, a key ingredient is the estimation of the void density profile. While previous studies have typically inferred this profile using a parameterized model based on simulations, we follow the approach of \cite{Boschetti} where the profile is directly extracted from the gravitational lensing signal of the voids. However, even though the transformation between the lensing signal and the density profile of dark matter in voids works effectively in \cite{Boschetti}, the voids definition adopted there differs from ours. Therefore, it is crucial to verify that this method remains valid for our void prescription. Similarly to \cite{Boschetti}, we use the Buzzard simulations \citep{derose1,derose2,wechsler, derose3} for which both the lensing signal for galaxies and the catalogues of dark matter particles are available. Our goal is to confirm that the  annular differential surface density measured from these two observables is consistent, and thus that

\begin{equation}
   \Upsilon_{\rm lensing}(r_{\rm p}/R_{\rm V},R_0)=\Upsilon_{\rm DM}(r_{\rm p}/R_{\rm V},R_0) \, .
\end{equation}

We identify voids in the galaxy catalogue and measure on the one side the lensing signal of such voids on background galaxies $\Upsilon_{\rm lensing}(r_{\rm p}/R_{\rm V},R_0) $ following Eq. \eqref{eq:ups1}, and on the other side, the density profile of voids in the particle field. We use the Davis\textendash Peebles \citep{davispeebles} estimator to measure the 2-point correlation function of each of our void samples with the dark matter particles in the simulations and a sample of random particles,

\begin{equation}
    \xi_{\rm Vm}(r_{\rm p}/R_{\rm V},\Pi)=\frac{N_{\rm Vm}(r_{\rm p}/R_{\rm V},\Pi)} {N_{\rm VR}(r_{\rm p}/R_{\rm V},\Pi)} -1\, ,
\end{equation}
where $N_{\rm Vm}(r_{\rm p}/R_{\rm V},\Pi) $ and $N_{\rm VR}(r_{\rm p}/R_{\rm V},\Pi)$ are the normalised number of void-galaxy or void-random pairs in each bin $(r_{\rm p}/R_{\rm V},\Pi)$. We then project this correlation function

\begin{equation}
    \Sigma(r_{\rm p}/R_{\rm V})=\bar{\rho}_{{\rm m}0} \int_{\Pi_{\mathrm{\min}}}^{\Pi_{\mathrm{\max}}}\xi_{\rm Vm}(r_{\rm p}/R_{\rm V},\Pi)\diff \Pi \, ,
\end{equation}
with $\bar{\rho}_{{\rm m}0}=\Omega_{\rm m}\rho_{\rm crit}$ and use the Eq. \eqref{eq:ups2} to compute $\Upsilon_{\rm DM}(r_{\rm p}/R_{\rm V},R_0)$. 

\begin{figure}
    \centering
    \includegraphics[width=1.0\linewidth]{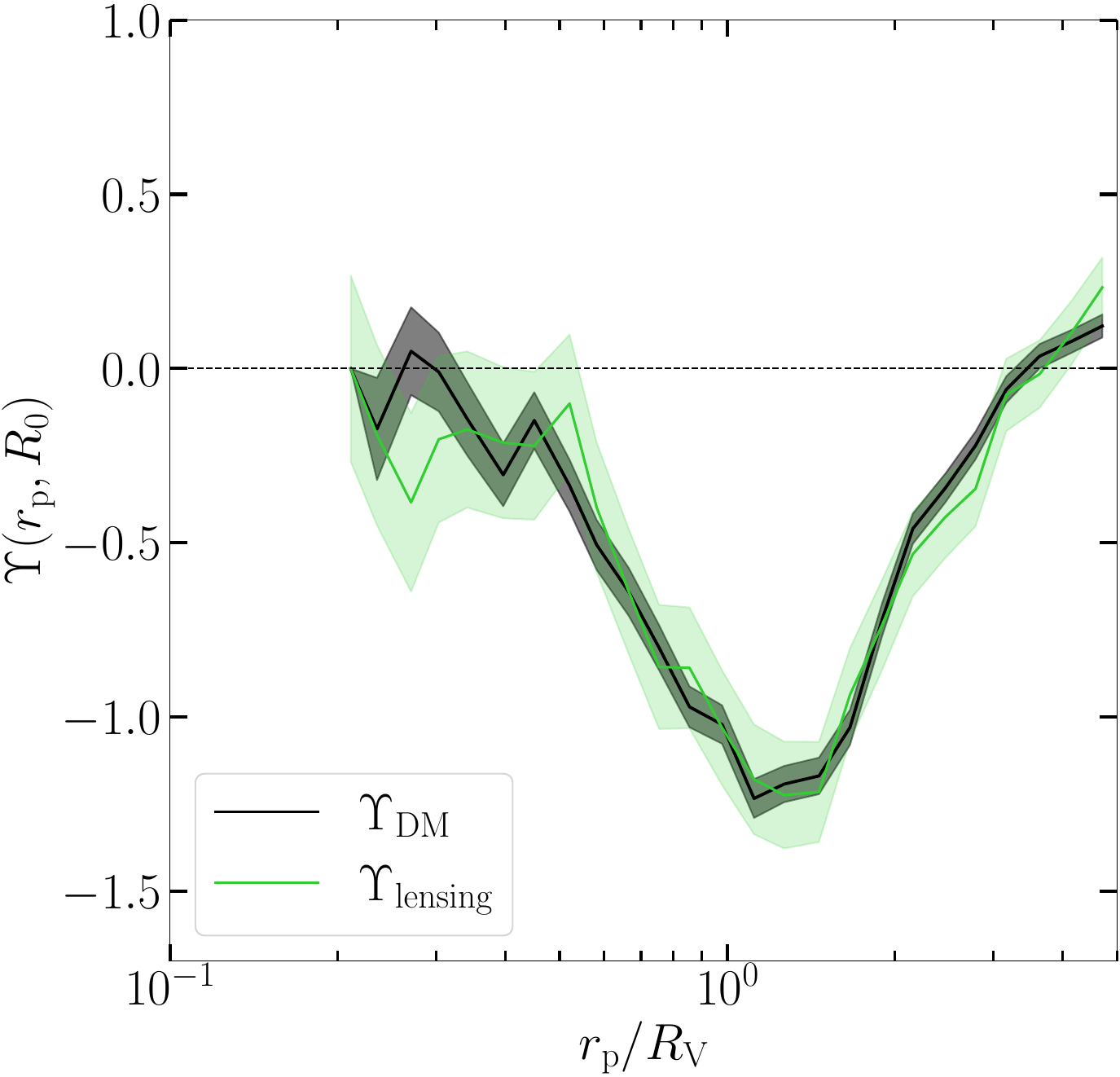}
    \caption {Comparison of the weak lensing signal of the voids $\Upsilon_{\rm lensing}$ to their profile in the dark matter $\Upsilon_{\rm DM}$. $1803$ voids have been found in the Buzzard simulation with a radius between $10$ and 15\,$h^{-1} {\rm Mpc}$ and a redshift between $0.1$ and $0.3$. The shaded areas represent the uncertainties measured using a jackknife method.}
    \label{fig:upsilon_compa1}
\end{figure}

\begin{figure}
    \centering
    \includegraphics[width=1.0\linewidth]{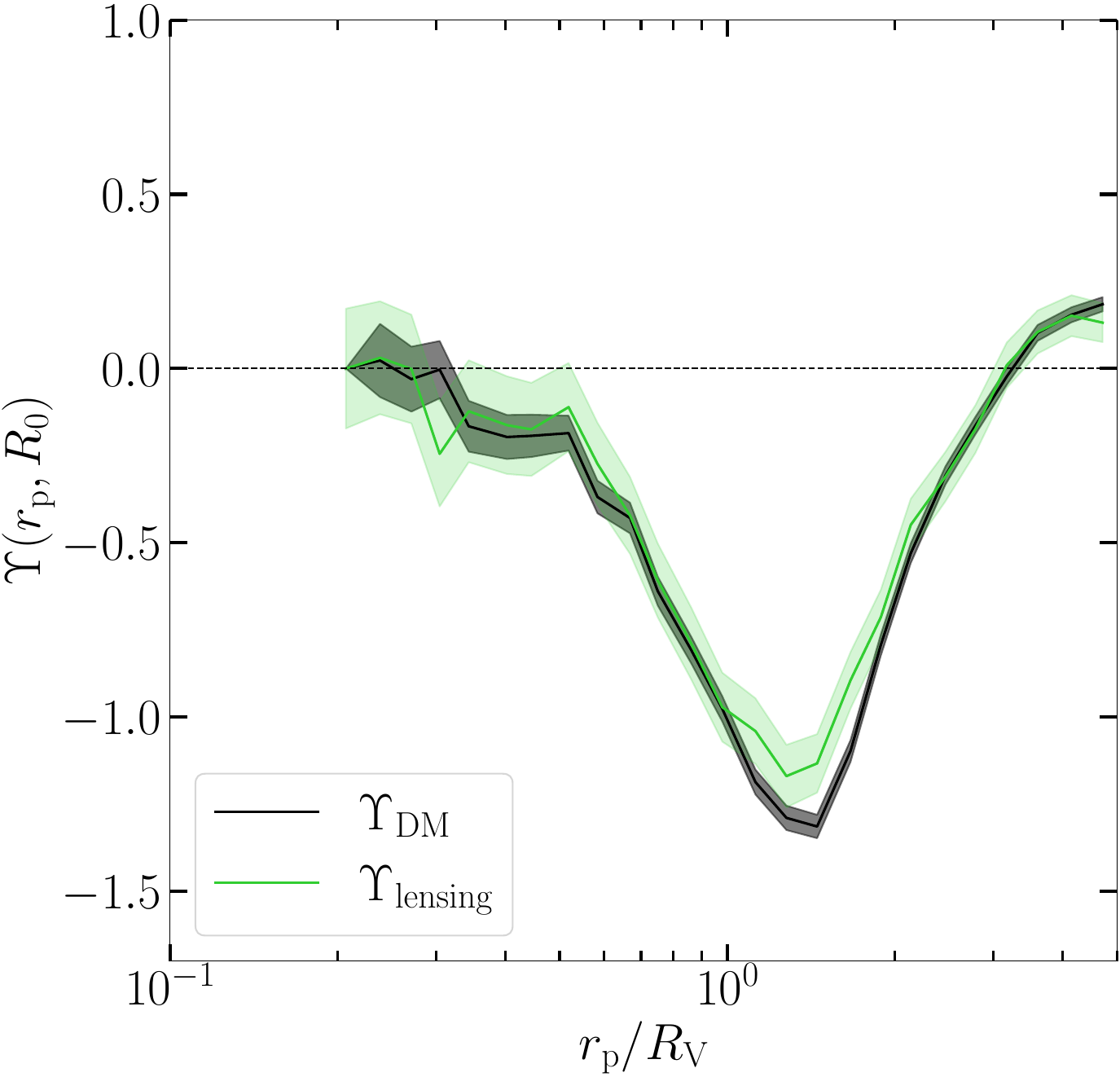}
    \caption{Same as Fig. \ref{fig:upsilon_compa1} for the redshift bin [$0.3$--$0.5$] (5438 voids). }
    \label{fig:upsilon_compa2}
\end{figure}

The good agreement between these two measurements of $\Upsilon $, for voids with a radius between 10 and 15\,$h^{-1} {\rm Mpc}$ and two bins in redshift, is shown in Figs. \ref{fig:upsilon_compa1} and \ref{fig:upsilon_compa2}.

\end{appendix}

\label{LastPage}

\end{document}